\documentclass[11pt]{article}

\usepackage[letterpaper,top=2cm,bottom=2cm,left=3cm,right=3cm,marginparwidth=1.75cm]{geometry}

\newcommand{\ARCjournal}{ARC Geophysical Research }
\newcommand{\ARCyear}{(2025) }
\newcommand{\ARCvolume}{1}
\newcommand{\ARCpaper}{---}
\newcommand{\ARCdoi}{doi:xxx}
\usepackage[utf8]{inputenc}
\usepackage{amsmath}
\usepackage{tabularx}
\usepackage{graphicx}
\newcolumntype{Y}{>{\centering\arraybackslash}X}
\usepackage{colortbl} 
\usepackage{xcolor}   
\usepackage{fancyhdr}
\fancypagestyle{plain}{\fancyhf{} \fancyhead[C]{{\scriptsize\color{blue} \ARCjournal \ARCyear \ARCvolume, \ARCpaper}} }
\fancypagestyle{fancy}{\fancyhf{} \fancyhead[L]{{\scriptsize\it \ARCauthors}} \fancyhead[R]{{\scriptsize\it \ARCjournal \ARCyear \ARCvolume, \ARCpaper}}\fancyfoot[C]{\thepage} }
\usepackage{authblk}

\usepackage[numbers, sort&compress]{natbib}

\usepackage[colorlinks=true, allcolors=blue]{hyperref}
\usepackage{orcidlink}

\usepackage{silence}
\usepackage{amsmath}

\usepackage{graphicx}
\usepackage{subcaption}

\usepackage{xcolor}

\newcommand\blfootnote[1]{%
  \begingroup
  \renewcommand\thefootnote{}\footnote{#1}%
  \addtocounter{footnote}{-1}%
  \endgroup
}

\usepackage[
    type={CC},
    modifier={by-nc},
    version={4.0},
]{doclicense}

\usepackage{lineno}
  
\usepackage{ctmmath-v3} 

\title{A stochastic modeling framework to generate 2-D rough-wall high-Reynolds-number turbulent boundary layers}

\author[1,2]{Roozbeh Ehsani\orcidlink{0009-0005-9273-6342}\thanks{Corresponding Author}}
\author[1,2]{Michele Guala\orcidlink{0000-0002-9788-8119}}

\affil[1]{{\scriptsize Saint Anthony Falls Laboratory, University of Minnesota, Minneapolis, MN 55414, USA}}
\affil[2]{{\scriptsize Department of Civil, Environmental, and Geo-Engineering, University of Minnesota, Minneapolis, MN
55455, USA}}

\date{}

\newcommand{\ARCauthors}{Ehsani et al.}

\begin{document}

\begin{figure}[ht!]
\centering
\includegraphics[width=\textwidth]{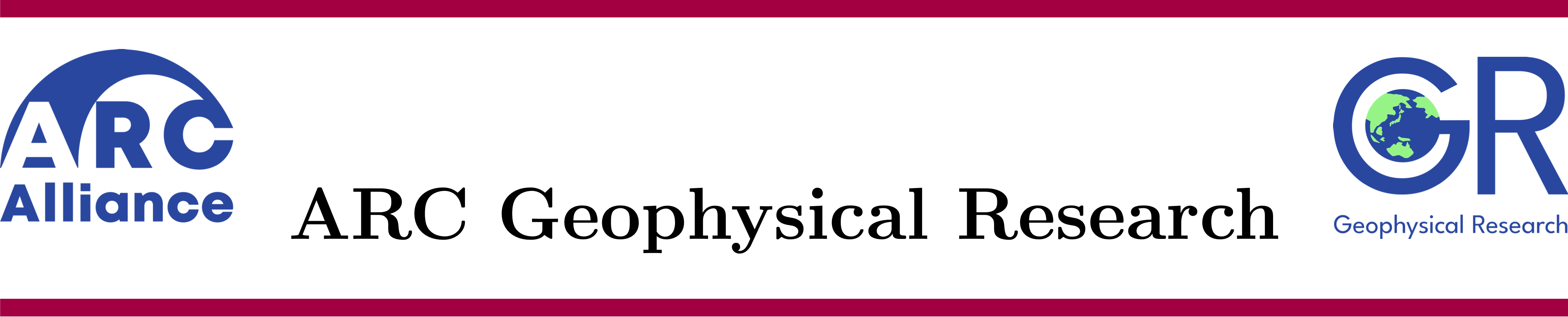}
\end{figure}
{\let\newpage\relax\maketitle}

\noindent
\hrule
\begin{abstract}

Atmospheric surface layer flows are computationally challenging, predominantly due to surface roughness and the high Reynolds number, both of which demand exceptionally high spatial resolution in the near-surface region. We have developed a 2-D stochastic-based model for the generation and the streamwise concatenation of instantaneous, step-like velocity profiles featuring key elements of wall turbulent flows, i.e., uniform momentum zones (UMZ) and shear layers \citep{ehsani2024stochasticprofile, ehsani2024stochasticfield}, and vortices \citep{ehsani2026}. The model is extended herein to the top of the logarithmic layer to reproduce the high-Reynolds-number, rough-wall, turbulent boundary layer measured by Saddoughi and Veeravalli \citep{saddoughi1994local}, without the support of a UMZ dataset, using only a handful of critical flow parameters:  the Taylor microscale $\lambda_{T}$, the boundary layer height $\delta$, the friction velocity $u_{\tau}$, and the aerodynamic roughness length $z_{0}$. The primary challenge lies in the integration of the stochastic model with the scaled distributions of the UMZ and vortex characteristics. The resulting statistical moments, energy spectra, and structure functions are compared against the experimental results. The validated code is made available in a GitHub repository.

\vskip  8 pt
\noindent
\hrule
\vskip 8 pt

\end{abstract}

\blfootnote{E-mail addresses: ehsan010@umn.edu}

\blfootnote{\ARCdoi}

\blfootnote{\vspace{-15pt}\doclicenseThis}

\vspace{-40 pt}

\section{Introduction}

\noindent Experimental studies of the near-neutral and convective atmospheric surface layer (ASL) near the wall within the logarithmic region indicate that turbulent structures across a wide range of length scales contribute substantially to Reynolds stresses $\tau_{ij}^{Re} = -\rho \overline{u_{i}^{\prime}u_{j}^{\prime}}$, heat, and vapor fluxes \citep{drobinski2004structure, bou2004large, nguyen2013measurements, heisel2018spatial, chung2021predicting, iungo2024grand}.
With recent advancements in numerical methods and computational power, large-eddy simulations (LES) have become an established tool for modeling the ASL and enabling predictive capabilities, thus providing a viable, affordable alternative to field experiments \citep{pascarelli2000multi,bou2004large,pantano2008approach, Moser23}. While wall-resolved large-eddy simulation (WRLES) provides high fidelity \citep{liu2025total}, its computational cost is, however, notoriously prohibitive for high-Reynolds-number applications; the required number of grid points scales as $\sim Re_{\tau}^{13/7}$ \citep{choi2012grid, lozano2019error}. As the Reynolds number increases, the separation between the Kolmogorov scale and the large and very-large scale motions, proportional to the boundary layer thickness $\delta$, expands significantly \citep{simens2009high,smits2011high,krank2016new}. This widening separation demands a progressively finer mesh to fully resolve small-scale turbulent structures, in particular within the near-wall region. To bypass the grid requirements of wall-resolved simulations, the strategy of under-resolving the near-wall region and utilizing wall functions to estimate the wall shear stress has been introduced \citep{deardorff1970numerical,schumann1975subgrid,kravchenko1996zonal, wang2002dynamic, schmidt2003near, bou2005scale, freire2021large, Yang24, zhou2024sensitivity, ghidoni2025assessment} and widely utilized. The first computational node adjacent to the wall in under-resolved LES of the ASL typically resides within the outer layer, necessitating the use of wall functions to prescribe the wall shear stress $\tau_{w}$ \citep{bose2018wall}. Consequently, any deficiency or structural limitation within the wall-modeling framework can lead to significant inaccuracies in the computation of near-wall forces, fluxes, and turbulence statistics in the simulation of the rough-wall turbulent boundary layer \citep{park2016wall, lozano2019error}. Because the outer flow relies on accurate turbulent boundary layer data, these uncertainties are not localized to the inner layer but rapidly propagate into higher elevations as structures grow and evolve in the logarithmic region, altering the overall boundary layer behavior and macroscopic characteristics \citep{bae2019dynamic, debroeyer2024wall}.

Ehsani {\it et al.} \cite{ehsani2024stochasticprofile} proposed a static, low-dimensional order, bottom-up stochastic model that utilizes uniform momentum zones (UMZs) as momentum-carrying structures to model the near-wall logarithmic region of a rough-wall turbulent boundary layer. UMZs carry a significant portion of the momentum fluxes and have been shown to represent the signature of the attached eddy structures proposed by Townsend \cite{townsend1976structure, adrian2000vortex, de2016uniform, heisel2020mixing, klein2024stochastic, winiarska2024flow, cui2025geometry, jing2025comparison, li2025characterization, kim2026uniform}; specifically, the thickness and modal (streamwise) velocity of the UMZs scale with the wall-normal distance $z$ and the friction velocity $u_{\tau}$, respectively. Each UMZ is characterized by three primary parameters: the zone thickness $h_{m}$, the modal velocity $u_{m}$, and the vertical velocity $w_{m}$. Identifying the correct distributions of UMZ attributes and employing UMZs' statistical properties along the wall-normal direction (mean $\mu(z)$ and standard deviation $\sigma(z)$), extracted from experimental Particle Image Velocimetry (PIV) \citep{heisel2020mixing} and super-large-scale particle image velocimetry (SLPIV) \citep{iungo2024grand} datasets, independent 1-D modal velocity profiles (step-like) were generated via the inverse transform sampling technique. The ensemble averages of the first- and second-order statistical moments of the generated profiles were observed to reproduce the logarithmic law of the wall (except in the immediate near-wall region) and yield acceptable levels of velocity variance and covariance \citep{ehsani2024stochasticprofile}. 

Building upon the generation of 1-D independent velocity profiles, this stochastic framework was subsequently extended by Ehsani {\it et al.} \cite{ehsani2024stochasticfield} to construct a correlated two-dimensional (2-D) velocity field by grouping similar step-like velocity profiles together. Spectral analysis of the stochastically generated velocity field demonstrated the reproduction of turbulence structures ranging from the energy-containing eddies to the inertial subrange, capturing the $k_{1}^{-5/3}$ regime. However, the smallest resolved length scale in that framework, defined by the spatial separation between adjacent velocity profiles, was bounded by the Taylor microscale $\lambda_{T}$. To resolve smaller energetic features beyond this limit, Ehsani {\it et al.} \cite{ehsani2026} refined the spatial resolution of the velocity field to introduce vortical structures into the 2-D domain. The study demonstrated that the introduction of vortices effectively injects kinetic energy into the inertial and dissipation subranges, thus extending the $k_{1}^{-5/3}$ regime, while preserving the $k_{1}^{-1}$ scaling characteristic of attached eddies. The height dependent distributions of vortex features, including the vortex core radius $r_{\omega}$, the maximum azimuthal velocity $u_{\omega}$, and a transformation parameter for the vortex velocity components, $\rho_{\omega}^{u,w}$, were derived from vortex detection on PIV data \cite{heisel2021prograde}. 

The scaling relationships developed to describe the probability density functions (p.d.f.) of the UMZs, shear layers, and vortex attributes require the friction velocity $u_{\tau}$, boundary layer thickness $\delta$, Taylor microscale $\lambda_{T}$, and aerodynamic roughness length $z_{0} \approx k_{s}/30$, where $k_{s}$ represents the equivalent sand-grain roughness. Therefore, given these four primary parameters for a target rough-wall turbulence flow field, this suite of stochastic models can generate a near-wall 2-D velocity field extending into the logarithmic layer and is able to satisfactorily reproduce both the moments of the flow and the spectral characteristics of the streamwise velocity component.  To validate the integration and test the performance of these models, the experimental high-Reynolds-number wall turbulent flow investigated by Saddoughi and Veeravalli \cite{saddoughi1994local} is reproduced here without a supporting PIV dataset, hence with no information on UMZs and vortex attributes' statistics and distributions. This challenge required several new features to be introduced in our model, which were designed to improve the reproduction of the energy spectra and the streamwise velocity variations across the entire logarithmic region. The long-term objective, which falls outside the scope of the present paper, is to reproduce the 3-D wall region of turbulent boundary layers with a stochastic architecture that could be coupled with LES to reduce the computational cost of ASL simulations, while preserving the genuine variability of near-wall turbulence. The following subsection briefly outlines the algorithms for the generation, concatenation, and spatial organization of the step-like velocity profiles, followed by the generation and seeding of the vortical structures into the modal flow field.

\subsection{Notation and Definitions}
\noindent To ensure clarity throughout the mathematical formulation of the stochastic model, the primary symbols, subscripts, and abbreviations are defined as follows:

\subsubsection*{Subscripts and Superscripts}
\begin{itemize}
    \item The subscript `$m$' indicates attributes belonging to the UMZs, including the zone thickness $h_{m}$, modal velocity $u_{m}$, and vertical velocity $w_{m}$. The bold notation $\mathbf{u}_{m} = (u_{m}, w_{m})$ represents the velocity vector containing both the streamwise and wall-normal components.
    \item The subscript `$\omega$' indicates attributes of an individual vortex, including the core radius $r_{\omega}$ and the maximum azimuthal velocity $u_{\omega}$.
    \item The subscripts `$i$' and `$j$' denote the index of a generated uniform momentum zone (UMZ) and vortex instance, respectively.
    \item The superscript `$^{\prime}$' denotes the fluctuating velocity component (e.g., $u^{\prime} = u - \overline{U}$), while the superscript `$+$' indicates normalization by inner scales ($u_{\tau}$ for velocity, $\nu/u_{\tau}$ for length).
\end{itemize}

\subsubsection*{Terminology}
\begin{itemize}
    \item The terms `characteristics', `features', and `attributes' are used interchangeably to describe the physical properties of UMZs and vortices.
    \item For the step-like velocity profiles, the streamwise velocity component ($u$) is represented interchangeably by the zone's modal velocity ($u_{m}$).
\end{itemize}

\subsection{Algorithm}
\noindent The stochastic generation of an organized 2-D rough-wall turbulent boundary layer velocity field, within the logarithmic region of the wall, featured by vortices seeded within the shear layers and near-wall regions, is achieved via a three-step procedure:
\\ 

\begin{figure}[ht!]
    \centering
    \includegraphics[width=1.0\textwidth]{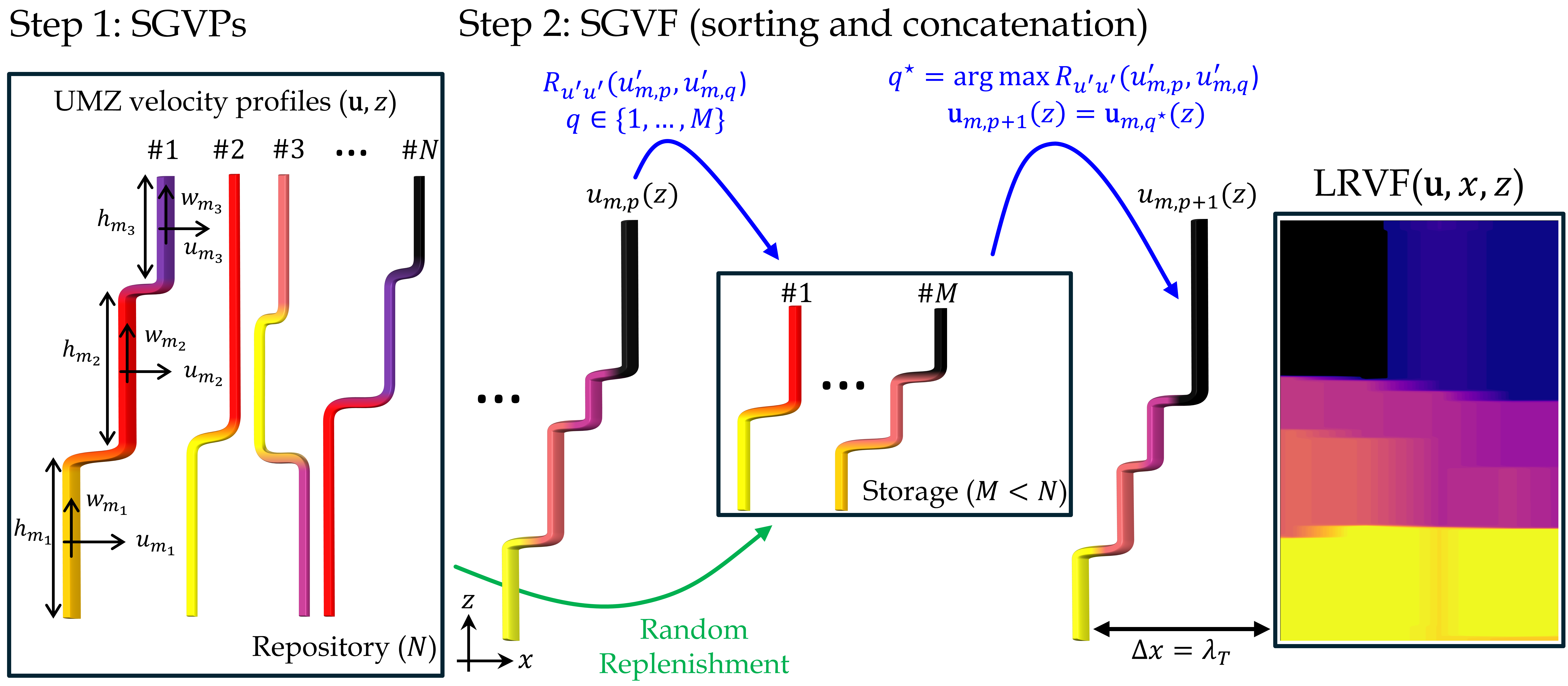}
    \caption{Schematics of the generation, sorting, and concatenation process. The size of the storage $M$ impacts the comparison of the cross-correlation coefficients between a given step-like velocity profile ($u_{m,p}$) and any other profile contained in $M$. Increasing $M$ leads to more similar profile concatenation and aggregation at larger scales. The most correlated velocity profile ($\mathbf{u}_{m,q^{\star}}$) is selected as the subsequent velocity profile ($\mathbf{u}_{m,p+1}$), spaced with the Taylor microscale ($\Delta x = \lambda_{T}$). The cross-correlation is computed using the fluctuating streamwise velocity component. Following each sampling operation from the storage buffer, the pool is randomly replenished from the profile repository to maintain a constant number of candidate velocity profiles.}
    \label{fig:schem}
\end{figure}

\textit{Step 1: Stochastic Generation of 1-D Step-Like Velocity Profiles (SGVP)}: To stochastically generate a step-like velocity profile, comprising multiple stacked UMZs, the generation sequence is initiated at a prescribed wall-normal elevation, $z_{\text{start}} = 1.5 k_{s}$, identified at the interface between the roughness sublayer and the logarithmic region where UMZ attributes can be collected. This process requires three primary parameters for each zone: thickness $h_{m_{\mathnormal{i}}}$, modal (streamwise) velocity $u_{m_{\mathnormal{i}}}$, and the vertical velocity $w_{m_{\mathnormal{i}}}$ of the UMZ. The governing equations and scaling laws for these variables are compiled in Tables~\ref{tab:equation} and \ref{tab:coeff} within appendix~\ref{app:equation}. To generate the $i$-th instance of these variables, three random numbers are sampled from a uniform distribution, such that \{$\text{rand}_{h_{m_{i}}}$, $\text{rand}_{u_{m_{i}}}$, $\text{rand}_{w_{m_{i}}}\} \sim \mathcal{U}(0, 1)$. To reproduce the Reynolds shear stress, the $\text{rand}_{u_{m_{i}}}$, and $\text{rand}_{w_{m_{i}}}$ are generated from a Gaussian copula with a prescribed correlation coefficient equal to $\rho_{m}^{\overline{u^{\prime}w^{\prime}}}$. This value can be estimated using an equation developed by Priyadarshana \cite{priyadarshana2004study} for high Reynolds numbers; further details are provided in section \ref{sec:pry}. Following the generation of the $i$-th UMZ, the subsequent zone is initialized at $z_{i+1} = z_{i} + h_{m_{i}}$, effectively populating the velocity profile in the wall-normal direction. This iterative sequence terminates once the wall-normal coordinate of the $(i+n)$-th zone, defined by $z_{i+n} = z_{i+n-1} + h_{m_{i+n-1}}$, exceeds the specified upper bound $z_{\text{end}} = 0.25\delta$. Any portion of a zone exceeding this limit is truncated, concluding the generation of a single independent 1-D velocity profile spanning from $z_{\text{start}} = 1.5k_{s}$ to $z_{\text{end}}=0.25\delta$. The generation of independent profiles continues until ensemble convergence is achieved and ensures the stationarity of statistical moments. A schematic of the $N$ generated 1-D step-like velocity profiles is illustrated within the repository block of Figure~\ref{fig:schem}, while a real sample from the generated velocity field is displayed in the inset~(1) of Figure~\ref{fig:VFcontour}(a). The comprehensive wall-normal statistical analysis of the UMZ characteristics and the underlying profile-generation framework are detailed in Ehsani {\it et al.} \cite{ehsani2024stochasticprofile}.
\\

\textit{Step 2: Stochastic Generation of a 2-D Velocity Field (SGVF)}: To construct a correlated 2-D velocity field, hereafter termed a low-resolution velocity field (LRVF), from the stochastically generated 1-D velocity profiles generated in Step 1, a sorting algorithm is implemented. Prior to sorting, to bound the velocity gradient at the UMZ interfaces, a smooth harmonic transition of modal and vertical velocity with a thickness of $\delta_{\omega} = 0.4\lambda_{T}$ is applied to the individual velocity profiles. The value of $0.4\lambda_{T}$ emerged as the average shear layer thickness across a wide range of Reynolds numbers \citep{de2016uniform, heisel2021prograde}. The total ensemble of generated profiles is aggregated into a primary repository, from which a subset of $M$ profiles is randomly sampled and transferred to a secondary storage buffer. The initial velocity profile of the 2-D velocity field is selected via random sampling from the repository. The adjacent downstream profile is obtained by scanning the storage buffer and selecting the candidate profile that exhibits the maximum cross-correlation of $u^{\prime}$ with its upstream neighbor. Once a profile is selected and placed into the 2-D domain, the storage buffer is replenished via random sampling from the repository to maintain a constant candidate pool size, $M$, for the next cross-correlation iteration and ensure the spatial homogeneity of the velocity variance. This iterative procedure is repeated consecutively for all subsequent downstream velocity profiles (i.e., $q^\star = \arg\max_{q \in \{1, \dots, M\}} R_{u^{\prime}u^{\prime}}(u^{\prime}_{m,p}, u^{\prime}_{m,q})$ where $p$ is the last profile in the velocity field, eventually $\mathbf{u}_{m,p+1}(z) = \mathbf{u}_{m,q^\star}(z)$). The buffer size $M$ is constrained such that the physical streamwise separation $\Delta x_{\text{LRVF}}$ between consecutive profiles matches the Taylor microscale $\lambda_{T}$. The Taylor microscale is assigned to the modal velocity field assuming vertical homogeneity to impose a regular grid. Technically, $\lambda_{T}=(15\nu\overline{u^{\prime 2}}/\epsilon)^{1/2}$, which can be simplified to $\lambda_T \propto (\nu/u_\tau)^{1/2} (kz)^{1/2} \sim \sqrt{l_{\nu} \ kz}$  assuming equilibrium between turbulence production $u_\tau^3/kz$ and dissipation $\epsilon$. The emerging mixed scaling between the viscous length scale ($l_\nu$) and a wall-attached length scale ($kz$) suggests a weak dependency on the elevation that could be overlooked in a relatively shallow layer where the above approximations may hold. In view of a low-dimensional modeling approach, it can be argued that the Taylor microscale could be estimated by reducing the strictly required parameters to $z_{0}$, $u_{\tau}$ and $\delta$. To eliminate initial transient effects and potential selection bias that impact the statistics, the first $\delta$-scale spatial segment of the reconstructed velocity field is discarded as a spin-up.

Because the vertical extent of the generated profiles is extended here to $z/\delta = 0.25$, as compared to $z/\delta = 0.15$ in Ehsani {\it et al.} \cite{ehsani2024stochasticprofile}, the storage size $M$ has to be adjusted to account for the enhanced variability of the velocity step profiles within the vertical domain. An investigation into the influence of storage size on energy production range is presented in Section \ref{sec:stor-size}. 
The schematic of the sorting and concatenation procedure is illustrated in Figure \ref{fig:schem}, and the foundational architecture for the construction of organized and Reynolds-number-dependent 2-D velocity fields is detailed in Ehsani {\it et al.} \cite{ehsani2024stochasticfield}.

\textit{Step3: Stochastic Generation of a Vortex (SGVorX)}: While the LRVF constructed in Step 2 successfully captures large-scale turbulent motions, its streamwise energy spectrum $E_{11}$ is fundamentally truncated at the Nyquist limit, resolving energy only down to the $L = 2\lambda_{T}$ scale. To resolve smaller, though still energetic sub-grid structures, the spatial resolution is refined via modified Akima (\textit{MAkima}) interpolation. This resolution enhancement yields a high-resolution velocity field (HRVF) with a grid spacing of $\Delta x_{\text{HRVF}} = \lambda_{T}/K_{r}$, where the refinement factor is set to $K_{r} = 10$ in the present study. The choice of the refinement factor $K_{r}$ is governed by a trade-off between computational cost and the option to resolve the dissipation length scale $\eta$.\\
Following the resolution enhancement, synthetic sub-grid vortices are superimposed onto the shear layers and within the immediate near-wall region. To construct the $j$-th synthetic vortex instance, parameterized by its core radius $r_{\omega_{j}}$, maximum azimuthal velocity $u_{\omega_{j}}$, and a structural transformation parameter $\rho_{\omega_{j}}^{u,w}$, three random numbers are drawn from a uniform distribution: $\{\text{rand}_{r_{\omega_{j}}}, \text{rand}_{u_{\omega_{j}}}, \text{rand}_{\rho_{\omega_{j}}^{u,w}}\} \sim \mathcal{U}(0, 1)$. The random variables $\text{rand}_{r_{\omega_{j}}}$ and $\text{rand}_{u_{\omega_{j}}}$ are drawn from a Gaussian copula with a prescribed Pearson correlation coefficient derived from experimental data: $0.4$ for $z_{j,c}/k_{s} < 2$ and $0.45$ for $z_{j,c}/k_{s} \ge 2$, where $z_{j,c}$ denotes the wall-normal elevation of the $j$-th vortex center. The governing scaling laws and their associated statistical coefficients are compiled in Tables \ref{tab:equation} and \ref{tab:coeff} in Appendix \ref{app:equation}.  The exact mathematical implementation of the underlying, transformed Oseen vortex model for the generation \cite{oseen1912uber}, alongside the superimposition procedure within shear layers and the near-wall region, is described in detail by Ehsani {\it et al.} \cite{ehsani2026}.
Following the superimposition of vortices within the shear layers and the near-wall region, a small-scale 2-D Gaussian filter with a characteristic width of $\Delta_{f} = 0.4\lambda_{T}$ is convolved over the velocity field. In the absence of an explicit model for viscous dissipation, the application of a convolutional filter mimics the effects of molecular viscosity on momentum diffusion and the transformation of turbulence energy into heat.

The fully stochastic framework formulated to generate the 2-D rough-wall turbulent boundary layer velocity field, encompassing profile generation to vortex superimposition, is governed by fundamental boundary layer parameters: the friction velocity $u_{\tau}$, the aerodynamic roughness length $z_{0}$, the boundary layer thickness $\delta$, and the Taylor microscale $\lambda_{T}$. In contrast to our previous data-driven approaches that rely on direct ingestion of experimental statistics during velocity field reconstruction, the present framework operates as a self-contained generative model. It requires these four parameters merely as baseline scaling inputs to independently synthesize the underlying uniform momentum zones and discrete vortex instances. The choice of the parameters emerges from their scaling relevance for the mean velocity profiles, the thickness of the shear layers and the velocity jump, the size of energetic vortex cores, and their azimuthal velocity \citep{heisel2021prograde, ehsani2026}. The fifth implicit parameter is the storage size $M$, which influences the large-scale motions \citep{ehsani2024stochasticfield}. 
For the validation cases presented in this study, these target flow parameters are prescribed based on the high-Reynolds-number wind tunnel measurements of Saddoughi and Veeravalli \cite{saddoughi1994local}. The remainder of this paper is organized as follows: \S~\ref{sec:modelparam} introduces the model parameters; \S~\ref{sec:enhancement} outlines modifications to the framework, including studies on storage size $M$, and the implementation of a band-stop filter; \S~\ref{sec:res} validates the resulting energy spectra and statistical moments against the benchmark dataset; and \S~\ref{sec:conclude} provides concluding remarks and future horizons for LES wall-modeling deployment.

\section{Model Parameters}\label{sec:modelparam}

\noindent To validate the model, the baseline parameters are chosen to match the high-speed boundary layer case documented by Saddoughi and Veeravalli \cite{saddoughi1994local}; these values are summarized in Table~\ref{tab:saddoughi}. Because the aerodynamic roughness length $z_{0}$ was not explicitly reported in the original experimental study, its value is determined herein by fitting the experimental mean velocity data within the logarithmic region ($5k_{s} \leq z \leq 0.20\delta$) to the law of the wall, assuming a Von K{\'a}rm{\'a}n constant of $\kappa = 0.39$.

\begin{table*}[ht!]
    \centering
    \caption{\label{tab:saddoughi} Flow and boundary parameters matching the high-speed wind tunnel dataset of Saddoughi and Veeravalli \cite{saddoughi1994local}. The reported friction velocity ($u_{\tau}$), boundary layer thickness ($\delta$), aerodynamic roughness length ($z_{0}$), and Taylor microscale ($\lambda_{T}$) serve as the primary scaling inputs to initialize the stochastic framework. The Taylor microscale is estimated using the statistics at $z/\delta = 0.36$.}
    \begin{tabular}{lccccccc}
        Dataset & $Re_{\tau}$ & $U_{\infty}[m/s]$ & $u_{\tau}[m/s]$  & $\delta[m]$ & $\lambda_{T}[mm]$ & $z_{0}[mm]$ & $\nu[m^2/s]$ 
        \\
        \cline{1-8} 
        \\
        High-Speed \cite{saddoughi1994local} & 16.8e4 & 50 & 2.32  & 1.09 & 7 & 0.38 & 1.5e-5  
        \\
        \cline{1-8}
    \end{tabular}
\end{table*}

\section{Model Integration and  Enhancements}\label{sec:enhancement}

\noindent This section details the structural modifications and novel physical features integrated into the present generative framework, and available in the attached code. These enhancements are systematically designed to improve the fidelity of the reconstructed turbulence statistics and energy spectra, while simultaneously automating the determination of internal model parameters through the coupling of well-established governing equations.

\subsection{Velocity Profile Generation Without Supporting UMZ Dataset}
\noindent The generation of the velocity step profiles requires the height-dependent statistics of UMZ attributes, e.g., $\mu(H_m(z)), \sigma(H_m(z))$. Ehsani {\it et al.} \citep{ehsani2024stochasticprofile} first fitted the required average and root-mean-square (rms) values, which were extracted from experimentally identified UMZs, and then proposed a set of dimensionless models for all the parameters as part of an a-posteriori analysis. Those functions, listed in Tables \ref{tab:equation} and \ref{tab:coeff}, were able to reasonably reproduce the wind tunnel and ASL UMZ statistics, but have never been used in the velocity profile generation process. The most relevant novelty of the present contribution is the implementation of the predictive models of height-dependent UMZ and vortex attributes for an untested flow condition without a supporting multi-dimensional UMZ and vortex datasets \citep{saddoughi1994local}, and the validation of the resulting modal velocity field.

\subsection{Extension to Height-Dependent Streamwise Velocity Variance}

\noindent 
To test the generated flow field across the whole logarithmic layer, we rely on the log law for the mean velocity profile. However, for the streamwise velocity variance, we need to adjust our model to account for the reduction of turbulent kinetic energy with $z$.
Invoking the attached eddy hypothesis, Townsend \cite{townsend1976structure} derived a generic logarithmic scaling law for the streamwise velocity variance $\overline{u^{\prime}u^{\prime}}$ within the logarithmic layer:

\begin{equation}
    \frac{\overline{u^{\prime}u^{\prime}}}{u_{\tau}^{2}} = -A\ln{(\frac{z}{\delta})}+B
    \label{eq:townsend}
\end{equation}

\noindent where $A$ and $B$ represent the Townsend-Perry constant and a flow-dependent empirical intercept, respectively. This scaling law is physically established by summing the contributions of the hierarchy of eddies, whose population scales inversely with the wall-normal distance ($\sim 1/z$), to the total energy spectrum. As shown by Qin \cite{qin2025asymptotic}, evaluations across multiple high-Reynolds-number datasets demonstrate that the Townsend-Perry constant, $A$, varies over a limited range between 0.9 and 1.33 \citep{perry1987experimental, nickels2007some, marusic2013logarithmic, samie2018fully, hwang2022logarithmic, huang2022profiles}. While the broad consensus in the literature frequently adopts a value of $A = 1.26$, which we used here, the intercept $B$ remains highly sensitive to the specific flow configuration. Based on the experimental statistics of Saddoughi and Veeravalli, and to compensate for the additional kinetic energy induced by the subsequent vortex superimposition step, a baseline value of $B = 1$ is prescribed in the present framework.

\subsection{Modeling the Streamwise-Vertical Velocity Correlation}\label{sec:pry}

\noindent Priyadarshana \cite{priyadarshana2004study} developed an empirical formulation relating the Reynolds shear stress correlation coefficient, $\rho^{\overline{u^{\prime}w^{\prime}}}$, to the momentum thickness Reynolds number, $Re_{\theta} = U_{\infty}\theta / \nu$, where $U_{\infty}$ denotes the free-stream velocity, $\Theta$ is the momentum thickness, and $\nu$ represents the kinematic viscosity. The value of $\Theta$ based on the boundary layer thickness can be approximated using the power law for the mean velocity profile (i.e., $\Theta \approx \frac{\delta n}{(n+1)(n+2)}$ where $n \approx 4$  for the rough-wall \citep{kotey2003power, dey2024approximate}). By introducing this scale mapping, the original formulation can be mathematically cast to establish a direct dependency on the friction Reynolds number, $\text{Re}_{\tau}$, as follows: $\rho^{\overline{u^{\prime}w^{\prime}}} = -0.03\ln{(\frac{0.15 U_{\infty}}{u_{\tau}}Re_{\tau})}+0.63$.\\
\indent While the Reynolds number dependency of the Reynolds shear stress correlation coefficient over smooth walls has been extensively demonstrated in the literature \citep{sillero2013one,morrill2015temporally,morrill2017reynolds}, it does not seem to converge on fully rough surfaces \citep{Morrill17}. Applying the proposed formulation yields a correlation coefficient of $\rho^{\overline{u^{\prime}w^{\prime}}} = -0.23$ for Saddoughi's 50 $[m/s]$ dataset, which underpredicts their experimentally reported value of $-0.33$. Given that this parameter is critical for the reproduction of the Reynolds shear stress, $\rho_{m}^{\overline{u^{\prime}w^{\prime}}} = -0.33$ is imposed in the stochastic generation of the modal and vertical velocity. Nevertheless, the algorithm retains the capability to employ Priyadarshana's formula \cite{priyadarshana2004study} when direct numerical or experimental estimates are unavailable.

\subsection{Storage Size Expansion}
\label{sec:stor-size}

\noindent The storage size $M$ required to construct a coherent spatial velocity field was originally determined empirically by extending the scaling range from the Taylor microscale defining the streamwise separation between consecutive 1-D velocity profiles and the resolved scale by UMZs ($\Delta x = \lambda_{T}$), to capture the largest flow structure, depending on $\delta$ and more broadly on the Reynolds number. By optimizing this parameter across multiple datasets \citep{heisel2020mixing,iungo2024grand}, Ehsani {\it et al.} \cite{ehsani2024stochasticfield} formulated a baseline scaling relationship of $M = 10\sqrt{Re_{\tau}}$. This empirical scaling aligns with the physical expectation that the extent of the inertial subrange broadens as the Reynolds number increases.\\ 
However, the spectral analyses of the streamwise velocity component identified an overestimation of energy within the inertial subrange \citep{ehsani2024stochasticfield}. This artifact is directly linked to the finite capacity of the storage utilized during the cross-correlation sorting routine. When $M$ is small, the restricted pool of candidate profiles causes a rapid downstream decay of spatial correlation, which mathematically manifests as an artificial inflation of the longitudinal structure function $D_{11}$ at inertial scales. A significant factor modulating the streamwise gradient of the $u_{m}$-velocity component, $\partial u_{m}/\partial x$, is the storage size utilized to impose a correlated sequence of vertical velocity profiles. To rectify this energy distribution while simultaneously accommodating an increased vertical domain size, extended from $z/\delta = 0.15$ in previous studies \cite{ehsani2026} to $z/\delta = 0.25$ herein, the present framework introduces a dual-modification strategy:

\begin{itemize}
    \item \textit{Storage Capacity Expansion}: The storage capacity $M$ is enlarged to allow for increased spatial correlations between step-velocity profiles extending further from the wall and thus possessing a higher degree of freedom.
    \item \textit{Spectral Notch Filtering}: A triangular band-stop filter is applied directly to the streamwise and cross-stream velocity signals to partially suppress the residual excess energy observed at the wavenumber scale corresponding to $k_{1} \sim 2\pi/\lambda$, as detailed in Section~\ref{sec:fourier-filter}.
\end{itemize}

 \indent While the Reynolds number dependency of storage size is documented, the impact of expanding the vertical extent of the generated profiles on horizontal field coherence has not been previously explored. Because an increased vertical height $h$ introduces greater structural variability among the synthesized profiles, the required storage capacity must scale as a coupled function of both parameters: $M = f(Re_{\tau}, h)$. Although an explicit analytical expression for this joint dependency is not derived in the current work, optimal values for $M$ were mapped empirically. Care must be taken while setting this parameter, as an indefinite increase of $M$ over-smooths the velocity field, yielding a gentle large-scale flow that lacks small-scale variability, a suppression of kinetic energy in the inertial subrange, and an overestimate in the production-range.

\begin{figure}[ht!]
    \centering
    \includegraphics[width=0.75\textwidth]{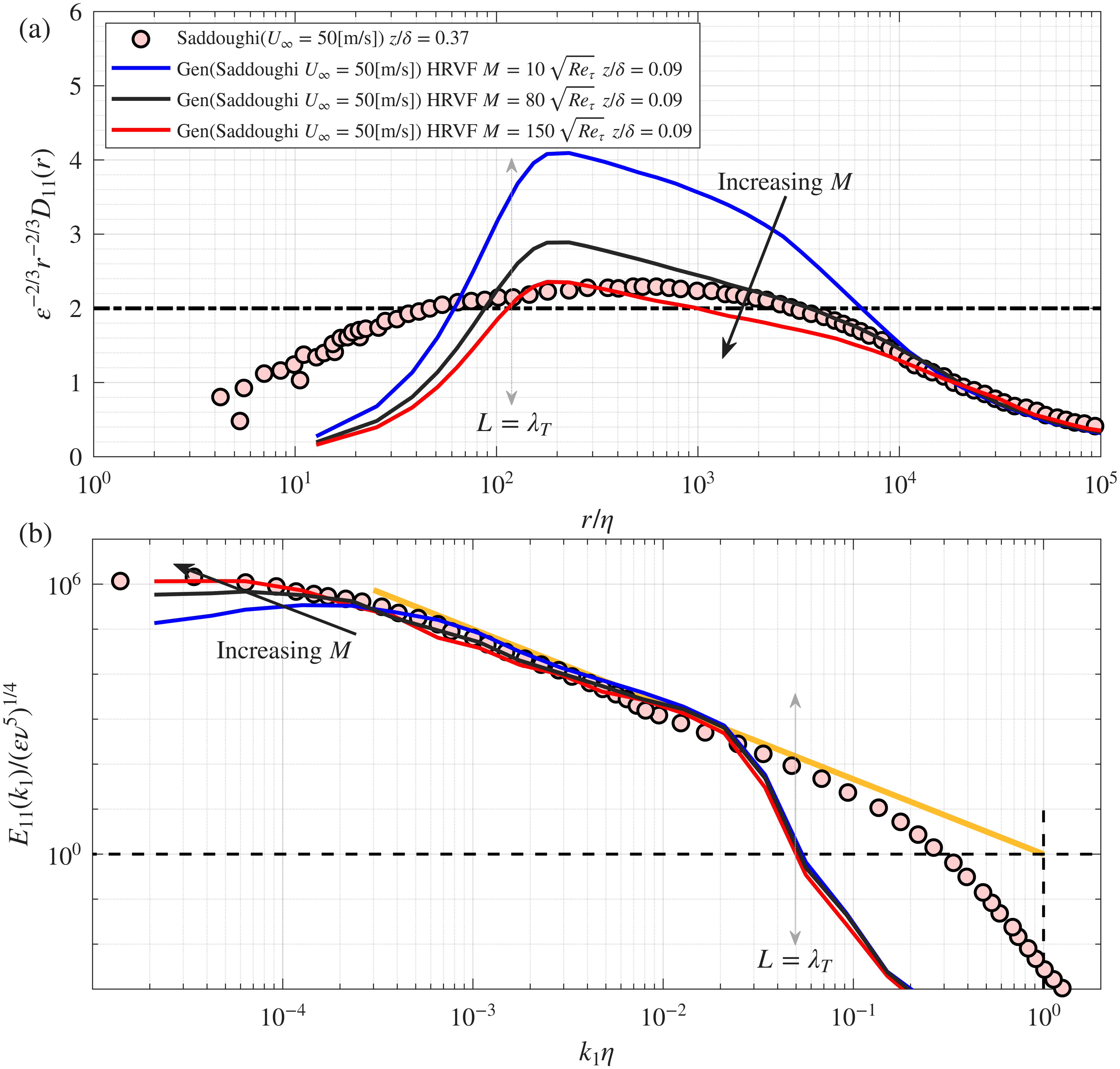}
    \caption{Influence of the storage capacity $M$ on the synthesized velocity field at $z/\delta = 0.09$: (a) longitudinal structure function $D_{11}$, and (b) energy spectra $E_{11}$ of the streamwise velocity component. The dissipation parameters ($\epsilon, \eta$) utilized for normalization are obtained from the reference experimental data \cite{saddoughi1994local}. Increasing $M$ systematically mitigates the overestimation of energy within the inertial subrange while shifting kinetic energy toward large production-scale modes.}
    \label{fig:storageeffect}
\end{figure}

\indent Figure \ref{fig:storageeffect} (a) illustrates the longitudinal structure function, $D_{11}$, computed across various storage sizes $M$. To systematically monitor the discrepancy between the synthesized fields and the reference dataset, the experimental energy dissipation rate, $\epsilon$, and the Kolmogorov length scale, $\eta = (\nu^{3}/\epsilon)^{1/4}$, are kept invariant while increasing $M$.  For other wall-normal elevations where Saddoughi and Veeravalli \cite{saddoughi1994local} did not explicitly report these parameters, $\epsilon$ is estimated using the local equilibrium production-dissipation relation, $\epsilon \approx u_{\tau}^3 / (\kappa z)$. At the specific validation height of $z/\delta = 0.09$, the values for the dissipation rate ($\epsilon$) and the Kolmogorov length scale ($\eta$) are reported as $342$ $[m^{2}/s^{3}]$ and $0.055[mm]$, respectively. 
In Figure \ref{fig:storageeffect}(a), we observe that an expansion of the storage size facilitates higher cross-correlation between sequential velocity profiles, effectively smoothing the local streamwise gradients of the $u_{m}$-component. This refinement serves to mitigate the overestimation of the structure function, $D_{11}$, in the inertial subrange, leading to a correct estimate of the dissipation rate and Kolmogorov scales. Concurrently, higher $M$ values strengthen the longitudinal correlation, sustain the large- and very-large-scale structures, and widen the inertial range (Figure \ref{fig:storageeffect}(b)). 

\subsection{Notch Filtering to reduce spatial resolution effects}
\label{sec:fourier-filter}

\noindent To establish the baseline 2-D low-resolution velocity field (LRVF), the stochastically generated 1-D profiles are concatenated horizontally with a uniform streamwise separation of $\Delta x_{\text{LRVF}} = \lambda_{T}$ \citep{ehsani2024stochasticfield}. Based on the spatial coherence and domain scale expansion criteria detailed in the previous section, the storage buffer capacity is fixed at an enlarged size of $M = 150\sqrt{Re_{\tau}}$. While this expanded buffer successfully preserves long-range spatial correlations, a residual overestimation of kinetic energy persists within the intermediate inertial-subrange scales, necessitating the implementation of a targeted spectral filter.  

\begin{figure}[ht!]
    \centering
    \includegraphics[width=0.75\textwidth]{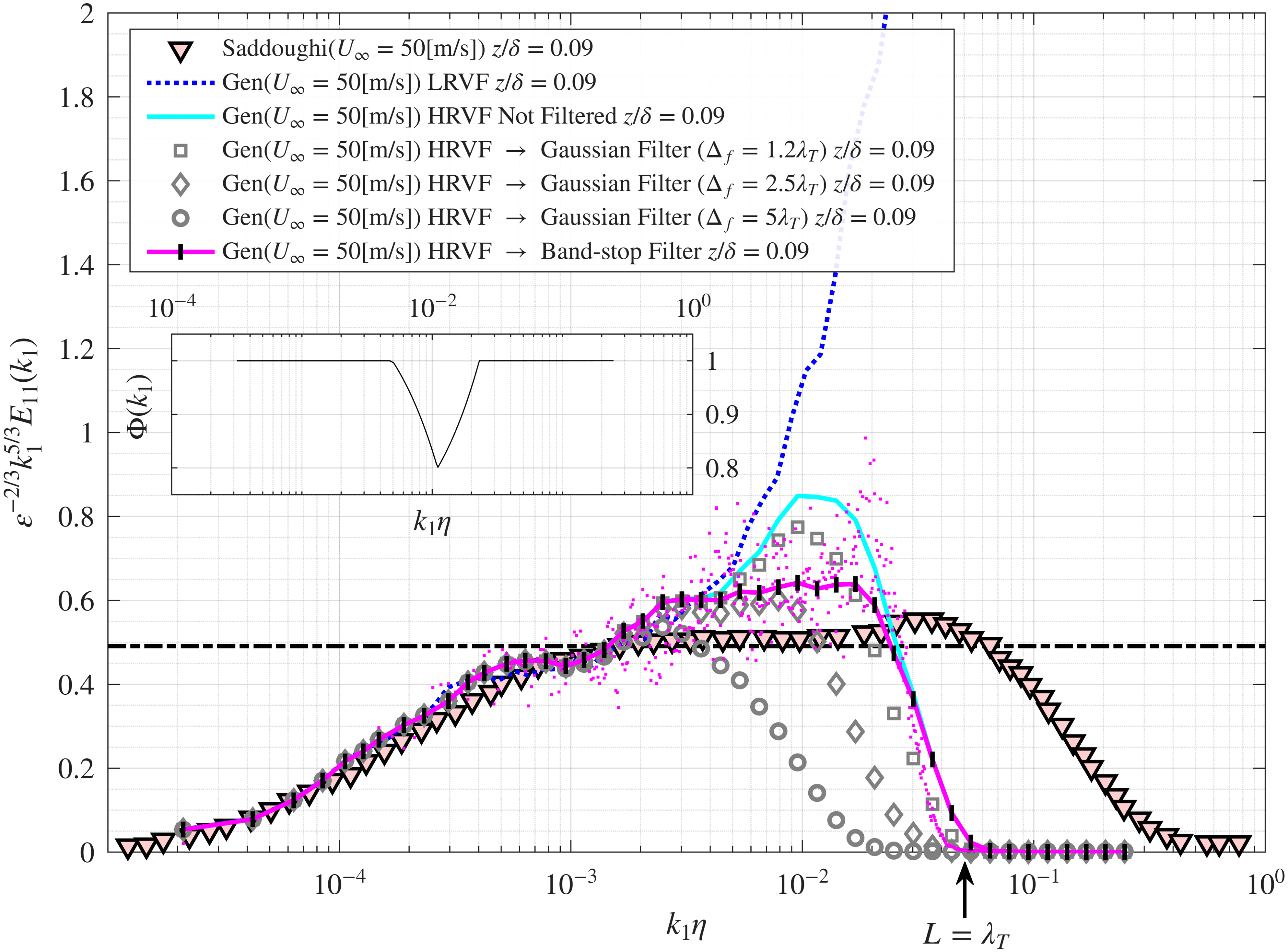}
    \caption{Pre-multiplied streamwise velocity energy spectra tracking the generative process from the low-resolution velocity field (LRVF) to the high-resolution velocity field (HRVF), evaluated against the reference experimental data of Saddoughi and Veeravalli \cite{saddoughi1994local}. The small-scale dissipation parameters ($\epsilon, \eta$) utilized for spectral normalization are derived directly from the experimental dataset. Comparative spectral profiles illustrate the impact of 1-D Gaussian filters at varying widths ($\Delta_f = 1.2\lambda_{T}$, $2.5\lambda_{T}$, and $5\lambda_{T}$) against the calibrated triangular notch filter. The Gaussian operators induce a broad, non-local attenuation across intermediate and high wavenumbers, whereas the targeted notch filter selectively suppresses energy strictly within the inertial subrange while preserving small-scale content. The filter maintains a unit amplitude at low wavenumbers before linearly declining at a corner wavenumber of $k_{1}\eta \approx 0.007\;(L \approx 10\lambda_{T})$. It reaches a minimum transmission magnitude of $0.8$ at $k_{1}\eta \approx 0.014\;(L \approx 5\lambda_{T})$, before recovering fully to a value of unity by $k_{1}\eta \approx 0.026\;(L \approx 2\lambda_{T})$. The wavenumber corresponding to the Taylor scale, $k_{\lambda_{T}} = 2\pi/\lambda_{T}$, is marked on the abscissa as a structural reference.}
    \label{fig:filtereffect}
\end{figure}

\indent Figure~\ref{fig:filtereffect} illustrates the pre-multiplied streamwise velocity energy spectra at key sequential stages of the generative process, tracing the structural evolution from the initial low-resolution velocity field (LRVF) to the high-resolution velocity field (HRVF). This spectral mapping explicitly assesses the efficacy of different filtering strategies in modulating the energy distribution across the resolved scale range. Because the LRVF resolves spatial structures down to the Taylor microscale ($\lambda_{T}$), its maximum resolvable wavenumber is bounded by the Nyquist limit, $k_{\text{max}} = \pi / \lambda_{T}$. For reference and comparison, the specific wavenumber corresponding to the Taylor scale, $k_{\lambda_{T}} = 2\pi/\lambda_{T}$, is explicitly marked on the abscissa. To enrich the sub-grid scales prior to vortex seeding, MAkima piecewise cubic Hermite interpolation is employed to refine the grid spatial resolution tenfold ($\Delta x_{\text{HRVF}} = \lambda_{T}/10$). This refinement factor represents a compromise between computational overhead and the requirement to keep the resolution to the order of the Kolmogorov length scale, $\eta$, to capture at least a portion of the dissipative range. Without filtering, the resulting high-resolution spectrum (``HRVF Not Filtered'') exhibits a distinct energy overestimation that peaks at $k_{1}\eta \approx 0.01$, corresponding to a length scale of $5\lambda_{T}$.

\indent To suppress this excess inertial-subrange energy, three distinct one-dimensional Gaussian filters, with characteristic widths of $\Delta_{f} = 1.2\lambda_{T}$, $2.5\lambda_{T}$, and $5\lambda_{T}$, are independently convoluted with the unfiltered $u_{m}$ velocity signal to systematically investigate the influence of filter scale on the spectral energy distribution. Because a standard Gaussian filter operates as a low-pass operator, it inherently attenuates energy across the entire high-wavenumber regime. Consequently, as the filter width broadens, this damping effect severely encroaches upon lower wavenumbers, leading to a non-local suppression of both total kinetic energy and larger-scale velocity variance.

\indent To specifically target energy attenuation within the intermediate inertial subrange while preserving the small-scale, high-wavenumber content, a triangular band-stop filter—operating as a spectral notch filter—is introduced. The transfer function characteristics of this filter are illustrated in the inset of Figure~\ref{fig:filtereffect}. The minimum attenuation coefficient of $0.8$ was established via empirical calibration against the reference spectra. In the wavenumber domain, the filtered velocity components are obtained as: $\hat{\mathbf{u}}_{\text{filtered}}(k{1}) = \hat{\mathbf{u}}_{\text{HRVF}}(k{1}) \cdot \Phi(k_{1})$, where $\hat{\mathbf{u}}$ denotes the Fourier-transformed velocity signal and $\Phi$ represents the triangular filter function. The modified signal is subsequently transformed back into physical space via an inverse Fast Fourier Transform (iFFT). The remaining localized energy deficit observed within the dissipation range is explicitly addressed through the subsequent introduction of discrete vortices, as elaborated in the next section~\ref{sec:res}. Note that the wave-number localization of all tested filters depends on $\lambda_T$ and thus is scalable to other flow conditions.

\section{Results}\label{sec:res}
\begin{figure}[ht!]
    \centering
    \includegraphics[width=0.75\textwidth]{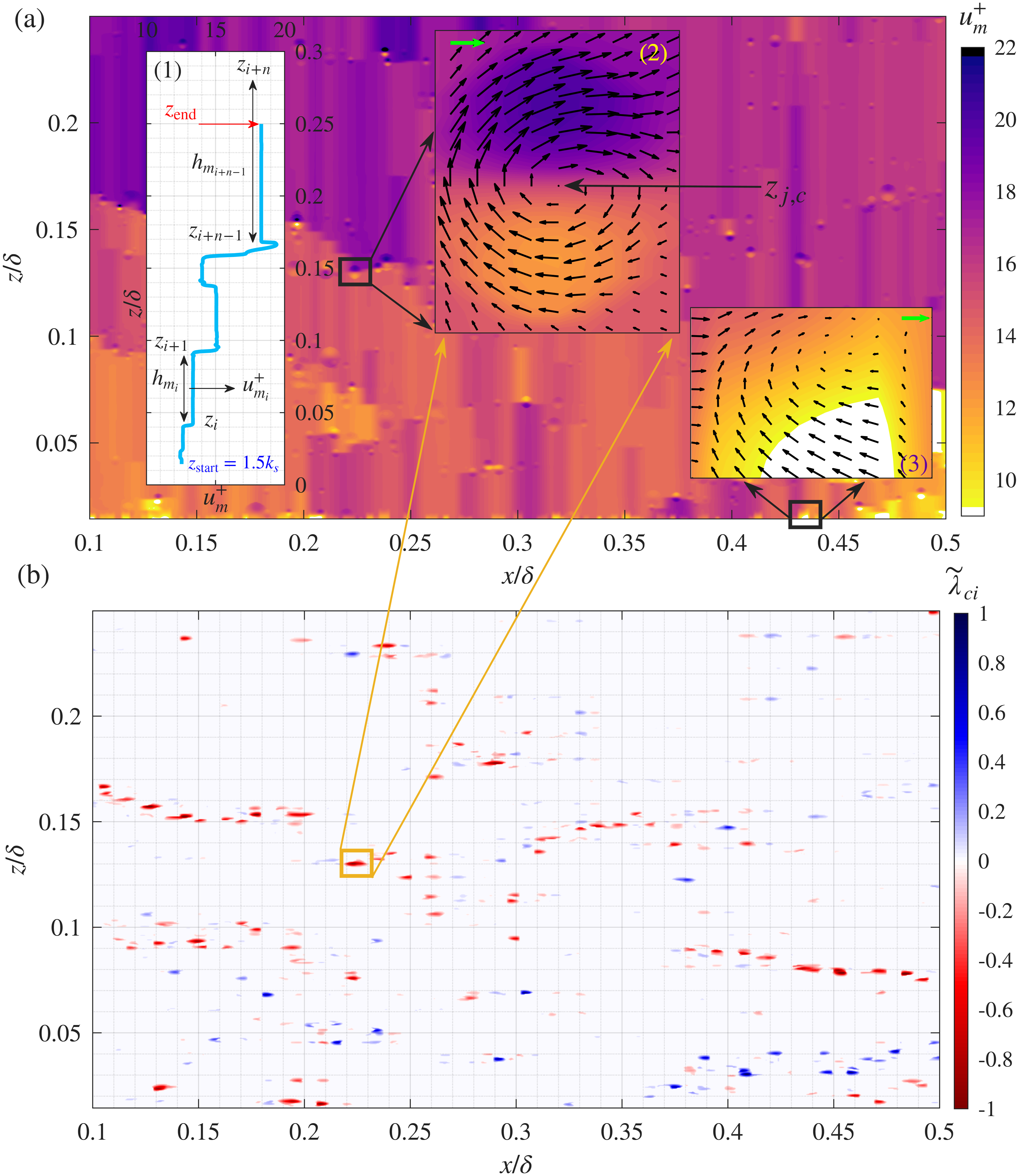}
    \caption{Spatial visualization of the fully synthesized velocity fields: (a) Contours of the normalized modal velocity component, $u_{m}^{+}$, for the high-resolution field following the superimposition of near-wall and shear layer vortices. Inset~(1) displays a localized slice of the step-like modal velocity profile at $x/\delta = 0.12$ spanning from $z_{\text{start}}$ to $z_{\text{end}}$. Inset~(2) shows the vector quiver of a representative prograde vortex within the shear layer interface; velocity vectors are rendered via a Galilean decomposition ($u_{m}^{+} - u_{\text{conv}}^{+}$) utilizing a local convection velocity of $u_{\text{conv}}^{+} = 15.43$ and a reference vector scale of $u_{\text{scale}}^{+} = 2$. Inset~(3) illustrates the vector quiver of a near-wall vortex, visualized using a convection frame of $u_{\text{conv}}^{+} = 11.6$. (b) Corresponding normalized signed swirling strength field, thresholded at $|\lambda_{ci}(x,z)| \ge 0.35\lambda_{ci,\text{rms}}(z)$ to isolate energetic turbulent structures. The scalar field is normalized by the maximum swirling strength of the domain, $\tilde{\lambda}_{ci}$, with the rotational sign prescribed by the spanwise vorticity component $\omega_{y}$.}
    \label{fig:VFcontour}
\end{figure}

\noindent The high-resolution velocity field is synthesized using a fixed storage capacity of $M = 150\sqrt{Re_{\tau}}$, the application of the triangular notch filter, and a population of shear layers and prograde near-wall vortices, followed by a final convolution with a small-scale 2-D Gaussian filter to model the physical effects of viscous momentum diffusion. Figure~\ref{fig:VFcontour}(a) depicts the contours of the $u_{m}-$ velocity component, normalized by the friction velocity. In the inset~(1), a representative slice of the step-like modal velocity profile at $x/\delta = 0.12$ is displayed. A spatially organized concatenation of these refined profiles yields the fully developed 2-D velocity field presented in Figure~\ref{fig:VFcontour}(a).

\indent The inset~(2) in Figure~\ref{fig:VFcontour}(a) presents a vector quiver plot of a representative prograde vortex seeded into a shear layer interface. To locate these shear layers, the local swirling strength field is computed as $\lambda_{ci} = |\Im(\lambda)|$, where $\lambda$ represents the complex conjugate eigenvalue pair of the velocity gradient tensor $\nabla\mathbf{u}_{m}$, as illustrated in Figure~\ref{fig:VFcontour}(b). The absolute imaginary component of this eigenvalue explicitly quantifies the local flow rotation. To distinguish the rotational direction, the signed swirling strength field is defined by the Hadamard product of $\lambda_{ci}$ and the sign of the local spanwise vorticity component, $\lambda_{ci} \odot \text{sgn}(\omega_{y})$, where $\omega_{y} = (\nabla \times \mathbf{u}_{m})_{y}$. Under this convention, a positive signed swirling strength corresponds to retrograde rotation, whereas a negative value indicates the placement of a prograde vortex. The discrete vortices are superimposed at the centroids of intense swirling strength clusters satisfying the threshold criterion $|\lambda_{ci}(x,z)| \ge \gamma \lambda_{ci,\text{rms}}(z)$, where $\lambda_{ci,\text{rms}}(z)$ is the wall-normal root-mean-square profile of the swirling strength, and the threshold coefficient is set to $\gamma = 0.35$. Adjusting $\gamma$ directly controls the population density of the small-scale structures; an excessively large value filters out higher-intensity rotational regions, thereby reducing the available seeding sites and artificially suppressing small-scale kinetic energy production. To fully populate each identified swirling strength cluster, a spatial constellation of secondary vortices is introduced into the shear layers using the iterative greedy algorithm detailed by Ehsani {\it et al.} \cite{ehsani2026}.

\indent The second family of structures introduced into the synthesized velocity field comprises prograde near-wall vortices, as illustrated in inset~(3). The wall-normal centers of these vortices are prescribed at a distance of $z_{j,c} = r_{\omega_{j}} + z_{\text{start}}$ above the wall, and they are stochastically seeded within regions of positive modal fluctuating velocity ($u_{m}^{\prime} > 0$) corresponding to sweep events. These high-speed sweeps statistically align with regions of intense Reynolds shear stress, which are characterized by high turbulent kinetic energy (TKE) and vorticity production \cite{hutchins2007large, guala2011interactions, iungo2024grand, wang2026}. Physically, this family of vortices populates the very first shear layer, dominated by roughness asperities and vortex shedding, and imposes a reduced slip condition at the lowest grid nodes to partially compensate for the finite, non-zero modal velocity at the wall. Full analytical expressions governing the velocity fields of both vortex families are provided in Ehsani {\it et al.} \cite{ehsani2026}. The subsequent sections evaluate the resulting turbulent statistics, spectral behavior, and spatial correlation topologies.

\subsection{Turbulence Statistics}
\label{res:turbstat}

\begin{figure}[ht!]
    \centering
    \includegraphics[width=0.7\textwidth]{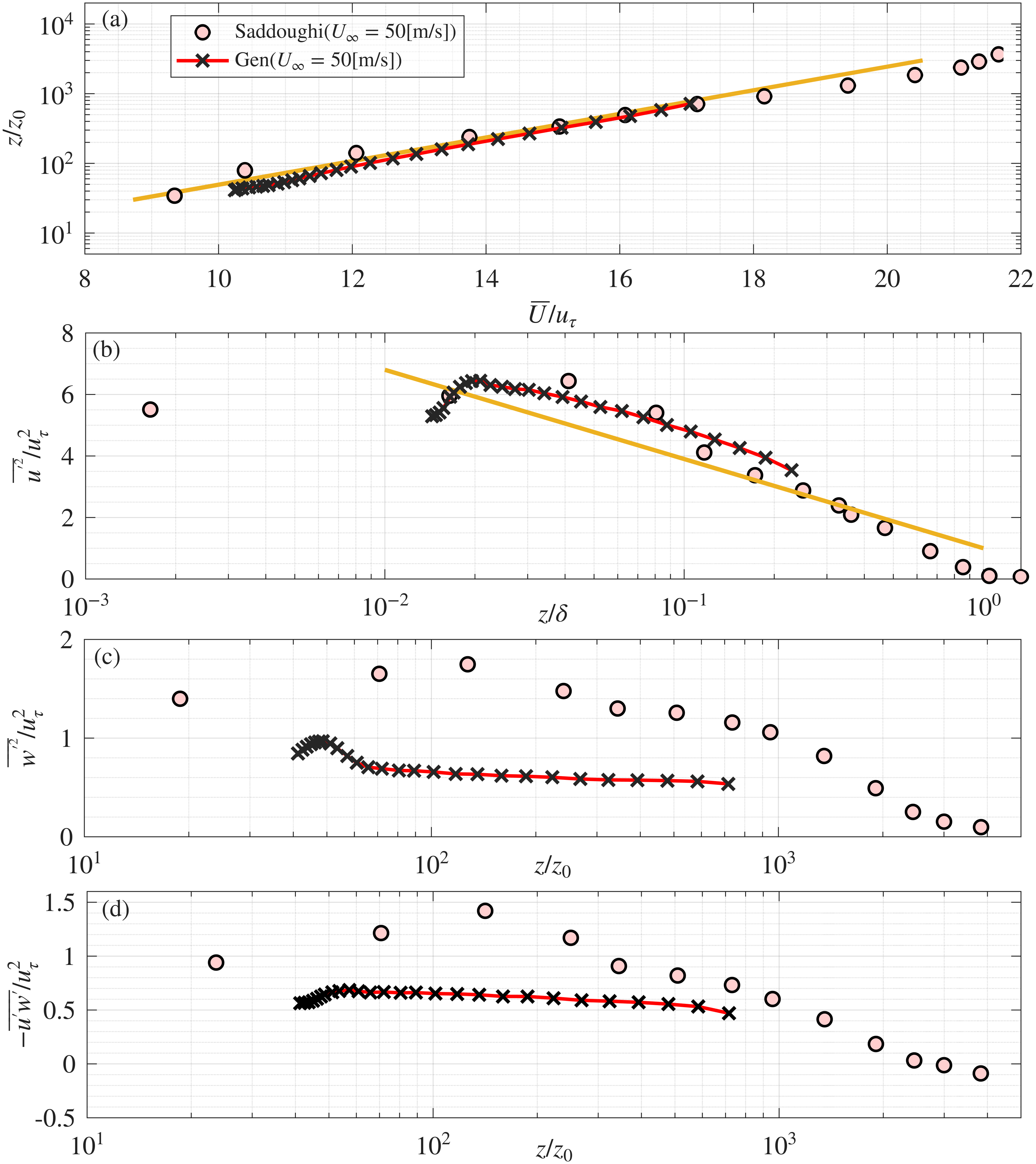}
    \caption{Comparison of the wall-normal profiles of the first- and second-order turbulent statistics for the synthesized velocity field against the reference experimental data of Saddoughi and Veeravalli \citep{saddoughi1994local}: (a) Mean streamwise velocity profile normalized by the friction velocity, $U^{+}$. The aerodynamic roughness length $z_{0}$ is determined by fitting the experimental data within the range $5k_{s} \leq z \leq 0.20\delta$ to the logarithmic law of the wall, $U^{+} = \kappa^{-1}\ln(z/z_0)$, using a von K{\'a}rm{\'a}n constant of $\kappa=0.39$. (b) Normalized streamwise velocity variance, $\overline{u^{\prime 2}}/u_{\tau}^2$. The theoretical baseline profile represents the Townsend attached-eddy formulation, $\overline{u^{\prime 2}}/u_{\tau}^2 = -A\ln(z/\delta)+B$, with calibrated parameters $A = 1.26$ and $B = 1$; the localized near-wall variance attenuation is induced by the superimposed prograde vortices. (c) Normalized wall-normal velocity variance, $\overline{w^{\prime 2}}/u_{\tau}^2$. (d) Normalized Reynolds shear stress profile, $-\overline{u^{\prime} w^{\prime}}/u_{\tau}^2$, generated via the coupled cross-correlation mechanisms $\rho_{m}^{\overline{u^{\prime}w^{\prime}}}$ within the uniform momentum zones and $\rho_{\omega}^{u,w}$ across the superimposed vortex populations, respectively.}
    \label{fig:stat}
\end{figure}

\noindent Figure~\ref{fig:stat} illustrates the wall-normal profiles of the first- and second-order turbulent moments for the synthesized velocity field evaluated against the reference experimental dataset. In Figure~\ref{fig:stat}(a), the theoretical logarithmic law of the wall is defined as $U^+ = \kappa^{-1}\ln(z/z_0)$. The mean streamwise velocity profile of the generated field exhibits a negligible overestimation in the immediate near-wall region. This discrepancy is attributed to the misdetection of UMZs within the high-shear near-wall region, which causes the omission of small-scale UMZs and their artificial coalescence into larger zones, thereby overestimating both the UMZ thickness statistics and the modal velocity \cite{ehsani2024stochasticprofile}. This discrepancy is significantly mitigated by the introduction of near-wall prograde vortices, as discussed above.

\indent The reduction of the normalized streamwise velocity variance $\overline{u^{\prime 2}}/u_{\tau}^2$ observed near the wall at elevations below $z/\delta \approx 0.02$, as shown in Figure~\ref{fig:stat}(b), is directly induced by the superimposed near-wall prograde vortices. Across the remainder of the wall-normal domain, the profile decays in parallel with the prescribed Townsend-Perry variance relation $\frac{\overline{u^{\prime 2}}}{u_{\tau}^{2}} = -1.26\ln(z/\delta) + 1$. Notably, the intercept computed from the synthesized velocity field shifts upward from this baseline, matching the experimental data more closely. This upward shift directly reflects the structural contribution of the embedded vortex populations. The logarithmic behavior of the variance reflects the cumulative contribution of wall-attached structures whose spatial extents scale with their distance from the wall, such that $\overline{u^{\prime 2}}/u_{\tau}^2 \propto \int_{z}^{\delta} z^{-1} \, \mathrm{d}z$ \cite{nikora1999origin}. This statistical signature is successfully reproduced by incorporating the wall-normal dependence of UMZ thickness in the modal velocity profiles, eventually leading to a correct velocity jump scaling with the friction velocity \cite{heisel2020mixing, ehsani2024stochasticprofile}, combined with the streamwise concatenation of correlated velocity profiles in the modal field reconstruction. The characteristic signature of wall-attached eddy structures is also manifest in the $k_{1}^{-1}$ scaling region of the streamwise energy spectrum $E_{11}$, which is equivalent to logarithmic scaling of the second-order structure function $D_{11}$, as discussed in the next section.\\
\indent The kinematic influence of the near-wall prograde vortices is also clearly manifested in the wall-normal velocity variance profile in the immediate vicinity of the wall, as shown in Figure~\ref{fig:stat}(c). A constant value of $\overline{w_{m}^{\prime 2}}^+ = 0.85^2$ is prescribed for the wall-normal modal velocity variance, consistent with the estimates of  \cite{ehsani2024stochasticprofile} at lower $Re_{\tau}$. Such a potential underestimation stems primarily from the vertical averaging of the wall-normal velocity within individual UMZs during the statistical extraction procedure \citep{ehsani2024stochasticprofile}.
We should however stress that in the logarithmic layer of a zero-pressure-gradient rough-wall boundary layer, the normalized wall-normal variance typically satisfies $\overline{w^{\prime 2}}^+ \approx 1.0 \text{--} 1.1$ \citep{Pope_2000, schlichting2016boundary}; hence the minor discrepancy observed between the synthesized vertical velocity variance and the experimental dataset may be also partly attributed to a potential overestimation of cross-stream fluctuations within the experimental measurements (or small underestimation of the the shear velocity). The $\overline{w_{m}^{\prime 2}}^+ $profile is assumed invariant across the remainder of the wall-normal domain (see Table~\ref{tab:coeff}), even though a vertical decay comparable with the streamwise velocity could be considered in future modeling efforts.

The observed underestimation in the vertical velocity fluctuations, as compared to the measured dataset, is similarly evident in the normalized Reynolds shear stress profile (Figure~\ref{fig:stat}(d)), which theoretically scales with the friction velocity such that $-\overline{u^{\prime}w^{\prime}}/u_{\tau}^2 \approx 1$. The primary mechanism governing Reynolds shear stress production in the present framework is the explicitly prescribed cross-correlation between the streamwise and the (reduced) vertical velocity components, $\rho_{m}^{\overline{u^{\prime}w^{\prime}}}$, across the generated UMZ steps, with a minor secondary contribution arising from the Pearson correlation coefficient, $\rho_{\omega}^{u,w}$, utilized during the transformation of the discrete vortex velocity fields. All those coefficients could be re-fitted, but would inevitably be inconsistent with the statistics of the UMZ attributes extracted from the \cite{heisel2020mixing, iungo2024grand} datasets.

\subsection{Spectral Analyses}

\begin{figure}[ht!]
    \centering
    \includegraphics[width=1\textwidth]{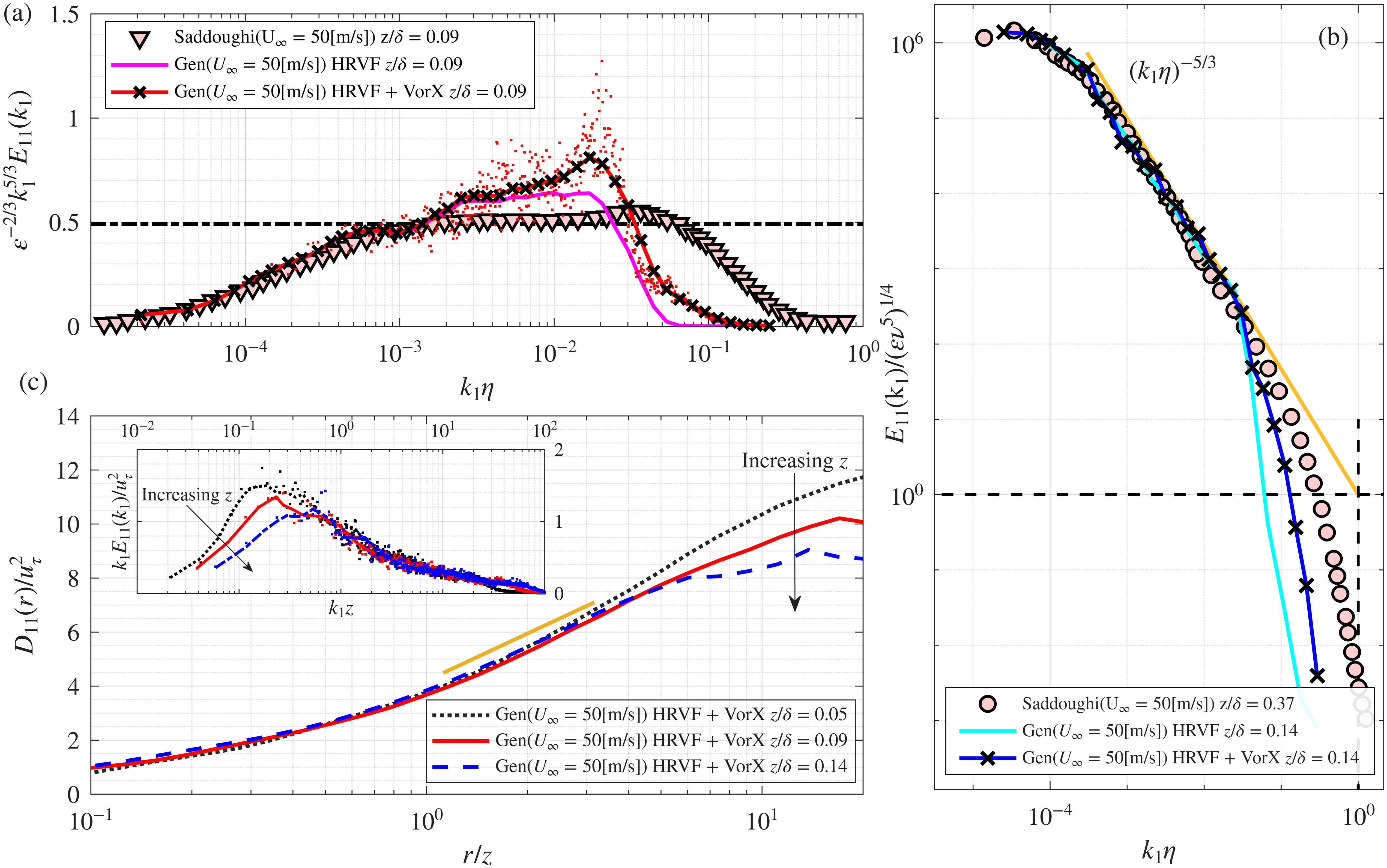}
    \caption{(a) Premultiplied streamwise velocity energy spectrum, $k_{1}^{5/3}E_{11}(k_{1})$, highlighting the signature of isotropic turbulence structure in the inertial subrange. The reference dissipation parameters ($\epsilon = 342~\mathrm{m^2/s^3}$ and $\eta = 0.055~\mathrm{mm}$) at $z/\delta = 0.09$ reported by Saddoughi and Veeravalli \cite{saddoughi1994local} are utilized to normalize the spectral curves for the generated velocity fields. The comparison between the baseline HRVF and vortex-populated (HRVF+VorX) fields highlights the systematic restoration of kinetic energy across the intermediate and dissipative subranges. (b) Energy spectrum $E_{11}(k_1)$ illustrating the distribution of energy across the complete cascade, spanning from the large-scale production structures down to the small-scale dissipative structures. The effect of superimposition of vortical structures is evident in the comparison between HRVF and HRVF+VorX. (c) Inner-normalized longitudinal second-order structure function, $D_{11}^{+}(r)$. The logarithmic scaling indicates the characteristic signature of wall-attached eddy structures, which is equivalent to $k_{1}^{-1}$ scaling of energy spectra shown in the inset plot. The theoretical yellow line follows $D_{11}^{+}(r) \sim 2A\ln{(r/z)}$, where $A=1.26$ is Townsend-Perry constant.}
    \label{fig:spectraD11}
\end{figure}

The premultiplied streamwise velocity energy spectra $k_{1}^{5/3}E_{11}(k_{1})$ are shown in Figure \ref{fig:spectraD11}(a), highlighting the correct signature of inertial range turbulence for the experimental dataset alongside the generated velocity fields before and after vortex superimposition. 
The effect of $\lambda_{T}$-scale vortex injection on the kinetic energy distribution is evident in the wavenumber range $10^{-2} \lessapprox k_{1}\eta \lessapprox 10^{-1}$, corresponding to the intermediate to dissipative scales.
Vortices increase the spectral estimate of the dissipation rate, $\epsilon = 15 \nu \int k_{1}^{2}E_{11}(k_{1})dk_{1}$, from $27[m^{2}/s^{3}]$ in the baseline HRVF model to $56[m^{2}/s^{3}]$ at a wall-normal location of $z/\delta = 0.09$ for the HRVF+VorX case.
Those values are, however, significantly smaller than the experimental value reported by Saddoughi and Veeravalli \cite{saddoughi1994local} ($\epsilon = 342[m^{2}/s^{3}]$ at $z/\delta = 0.09$), which is expected due to the lower spatio-temporal resolution and the quadratic weighting of $k_{1}$ in the dissipation integral, which amplifies the impact of under-resolved small-scale energy.
To correctly estimate the dissipation rate from all synthetic velocity fields, it is advantageous to use the compensated second-order structure function $D_{11}(r)$, and the emerging inertial-range plateau imposing $\epsilon = [(2.3)^{-1}\times \max \{ r^{-2/3} D_{11}(r) \}]^{3/2}$. Employing this method yields an estimated dissipation rate of $\epsilon = 352 [m^{2}/s^{3}]$, which exhibits much closer agreement with the benchmark value reported by experiment and confirms the correct reproduction of turbulence in the inertial range (see also the ${-5/3}$ slope in Figure \ref{fig:spectraD11} (b).

To further investigate the asymptotic behavior of the resulting synthetic velocity fields, we now focus on the other region of the streamwise velocity spectra. At high-Reynolds-numbers, there exists a specific range of wavenumbers located between the production subrange and the inertial subrange, bounded by $z < k_{1}^{-1} \ll \delta$, where the streamwise energy spectrum $E_{11}(k_{1})$ exhibits a characteristic $k_{1}^{-1}$ scaling \cite{nikora1999origin}. This specific turbulent kinetic energy distribution is associated with a hierarchy of wall-attached eddy structures, spanning length scales from $z$ up to a fraction of $\delta$, consistent with Townsend's hypothesis and responsible for the logarithmic decay of the streamwise velocity variance in the overlap region \citep{davidson2006logarithmic}. As the wall-normal elevation $z$ increases, the separation between the large-scale boundary layer motions and $z$-scale motions diminishes, thereby narrowing the wavenumber range over which the $k_{1}^{-1}$ scaling can be discerned. Furthermore, due to the artificial accumulation of energy at low wavenumbers arising from spectral aliasing, the signature of the wall-attached eddy structures emerges more clearly from the analysis of the second-order structure function $D_{11}(r)$ in physical space rather than in wavenumber space \cite{tennekes1972first, davidson2006logarithmic}. Figure \ref{fig:spectraD11}(c) illustrates the inner-normalized longitudinal second-order structure function. The signature of the wall-attached eddy structures is corroborated by the logarithmic behavior of the structure function, $D_{11}^{+}(r) \sim 2A\ln(r/z)$, within the spatial separation range $1 < r/z \ll \delta/z$, where $A = 1.26$ represents the Townsend-Perry constant. The complete derivation of this relationship, transforming the $k_{1}^{-1}$ scaling in the energy spectrum $E_{11}(k_1)$ to the structure function $D_{11}(r)$, is provided by Davidson \cite{davidson2015turbulence}.  This scaling is further validated by the pre-multiplied spectral energy plateau, $k_{1}E_{11}(k_{1})$, shown in the inset, which remains aligned with the prescribed value of $A = 1.26$. As the spatial separation approaches the boundary layer scale ($r \sim \delta$), the structure function plateaus toward its asymptotic limit, $D_{11}^{+}(r \sim \delta) = 2\overline{u^{\prime 2}}/u_{\tau}^2$, which decreases monotonically with increasing wall-normal distance in strict accordance with attached-eddy constraints \cite{townsend1976structure}. This is a strict validation check on the quality of our synthetic velocity field, across a wide range of scales.\\

\begin{figure}[ht!]
    \centering
    \includegraphics[width=0.7\textwidth]{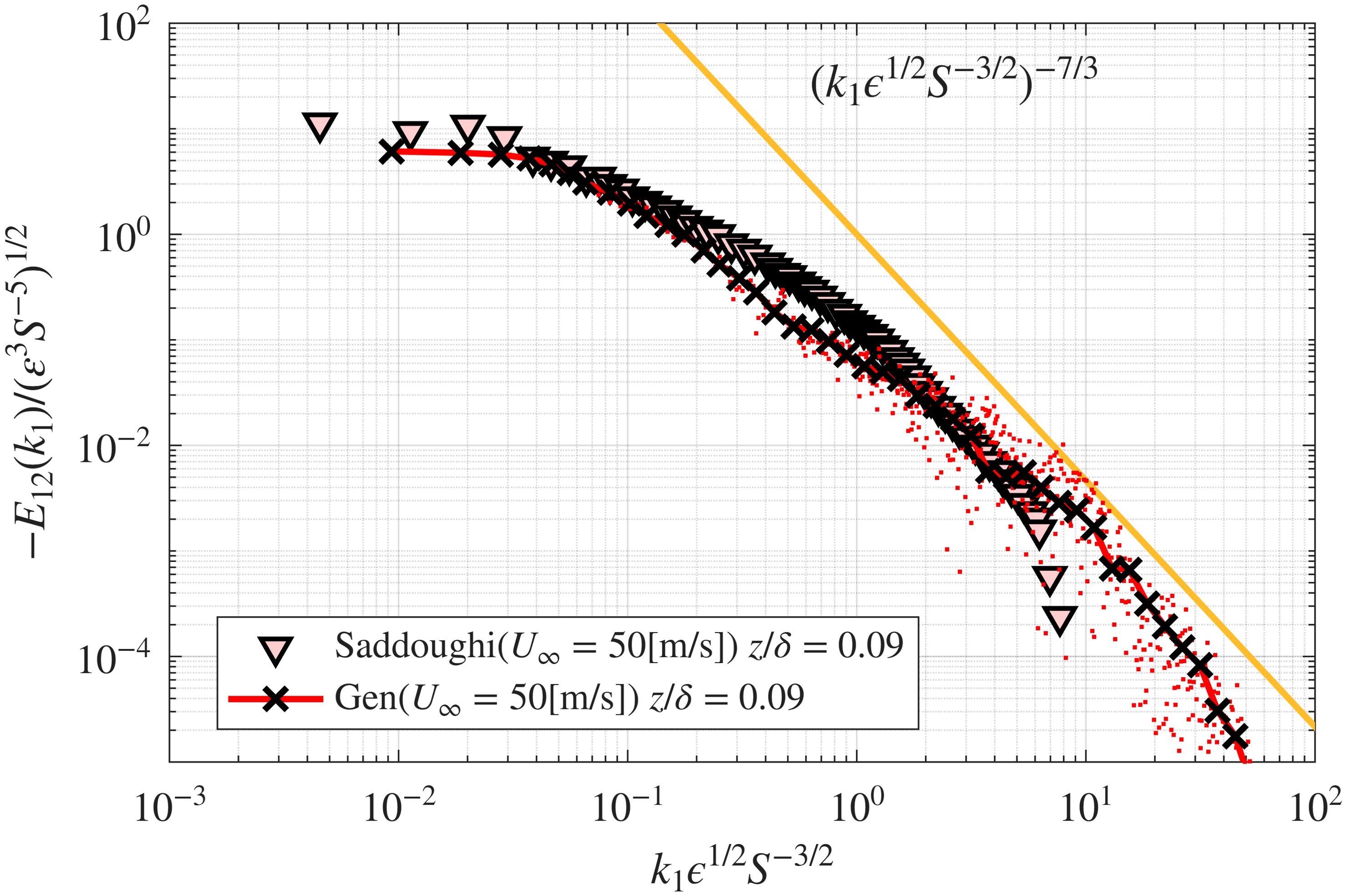}
    \caption{Turbulent shear-stress cospectrum, $-E_{12}(k_1)$, of the synthesized velocity field evaluated against the reference experimental measurements of Saddoughi and Veeravalli \cite{saddoughi1994local}. The curves are normalized utilizing the shear and dissipative scaling parameters derived from the experimental dataset ($S = 83.5~\mathrm{s^{-1}}$ and $\epsilon = 342~\mathrm{m^{2}/s^{3}}$). Based on these parameters, the theoretical inertial subrange is mapped within the bounds $0.44 \le k_{1}\epsilon^{1/2}S^{-3/2} \le 22$, demonstrating the extended scaling capacity of the synthesized field relative to the experimental dataset.}
    \label{fig:E12}
\end{figure}

\indent Figure~\ref{fig:E12} illustrates the turbulent shear-stress co-spectrum, $-E_{12}(k_{1})$, of the synthesized velocity field evaluated against the reference experimental dataset. In wall-bounded flows, the mean velocity gradient ($S = \partial U/\partial z$) sustains the production of turbulent kinetic energy, leveraging the cross-correlation between the streamwise and vertical velocity components, which govern the Reynolds shear stresses, vertical turbulent fluxes, mixing and dispersion, and other near-surface mechanisms. Within the inertial subrange, the $uw-$cospectrum depends explicitly on both the energy dissipation rate $\epsilon$ and the localized mean shear $S$. Dimensional analysis within this shear-dominated regime yields a characteristic $-7/3$ scaling law for the $uw-$cospectrum \cite{ganapathisubramani2003characteristics, guala2006large, balakumar2007large}. 

\noindent Given that the inertial subrange resides within the normalized wavenumber band $0.001 \lessapprox k_{1}\eta \lessapprox 0.05$ based on the energy spectra in Figure~\ref{fig:spectraD11}(a), the corresponding interval mapped to the inner-shear normalized abscissa, $k_{1}\epsilon^{1/2}S^{-3/2}$, spans from $0.44$ to $22$. The experimental dataset captures the theoretical $-7/3$ scaling up to $k_{1}\epsilon^{1/2}S^{-3/2} \approx 3$, while the synthesized velocity field extends the scaling region to the end of the inertial range. The localized underestimation of the cospectrum is observed at lower wavenumbers. $k_{1}\epsilon^{1/2}S^{-3/2} \approx 0.4$ at the opposite side of the inertial range stems directly from a structural deficit in the vertical velocity energy spectrum, $E_{22}$, within the same range. This is likely induced by i) the streamwise velocity-driven (solely $u$-based) organization algorithm employed during the velocity profile concatenation, and ii) the underestimation of the vertical modal velocity variance resulting from the averaging operator across each UMZ thickness, acknowledged above and amply discussed in \cite{ehsani2024stochasticprofile}.

\section{Discussion}\label{sec:discustion}

\noindent In this section, we evaluate the operational viability and architectural constraints of the proposed predictive stochastic framework. To assess its feasibility as a synthetic turbulence generator, the analysis is partitioned into two core components. First, we investigate the computational efficiency and memory footprints of the algorithm. Second, we delineate the inherent physical and structural limitations currently facing the model framework, providing a critical assessment.

\subsection{Computational Cost}

\noindent To conservatively evaluate the overall computational cost of the proposed stochastic framework, the velocity field generation pipeline is decomposed into its core sequential operations. The total execution time and resource allocation are primarily governed by the following operational phases:\\

\begin{itemize}
    \item \textbf{Profile Generation (SGVPs):} 
    To initialize the model, random numbers are first generated for the UMZs' characteristics, requiring a baseline time and space complexity that is negligible compared to the subsequent profile generation phase. The Space complexity for the generated profiles depends on the total number of generated profiles in the repository $(N_{\text{Prof}})$, the wall-normal spatial extent $[z_{\text{start}}, z_{\text{end}}]$, and the vertical grid resolution $\Delta z$. In this framework, the grid spacing is set to $\Delta z = 0.06 \lambda_{T}$ to resolve the internal shear layers with high fidelity, yielding a discrete number of wall-normal grid points $N_{z}$. Consequently, the total space complexity of this phase scales as $\mathcal{O}(N_{\text{prof}} \cdot N_{z})$. Note that $N_{\text{prof}}$ contains both velocity components. The corresponding time complexity scales as $\mathcal{O}(N_{\text{prof}} \cdot \langle N_{\text{steps}} \rangle)$, where $\langle N_{\text{steps}} \rangle$ represents the average number of discrete momentum zone steps per profile, which is governed directly by the targeted wall-normal domain height.
    
    \item \textbf{Smoothing the Shear Layers:} Prior to profile sorting and concatenation, the discrete transitions of the modal and vertical velocity components between adjacent UMZs at shear layers are smoothed utilizing harmonic functions. The average number of internal shear layers embedded per profile can be approximated by the average number of steps (i.e., $\langle N_{\text{SL}} \rangle \approx \langle N_{\text{steps}} \rangle $). The time complexity of this phase scales as $\mathcal{O}(N_{\text{prof}} \cdot \langle N_{\text{steps}} \rangle \cdot N_{z})$. Because this execution is performed in-place directly on the existing array allocations of the generated profiles, the auxiliary space complexity of this operation is bounded at $\mathcal{O}(1)$.
    
    \item \textbf{Sorting and Concatenation (SGVF-LR):} Although this operation is highly optimized via vectorization, the primary computational bottleneck of the model resides within this phase, where consecutive step-like velocity profiles are sorted and concatenated to construct a spatially correlated 2-D velocity field. The time complexity of this step scales approximately as $\mathcal{O}(N_{\text{prof}} \cdot M \cdot N_{z})$. Note that the final number of structured velocity profiles embedded within the generated spatial domain is approximated by $N_{\text{prof}}$. The corresponding space complexity scales as $\mathcal{O}((N_{\text{prof}}) N_{z})$.

    \item \textbf{Increasing the Resolution (SGVF-HR):} Before incorporating subgrid-scale structures into the low-resolution velocity field (LRVF), the spatial resolution is increased utilizing cubic Hermite interpolation. Because this interpolation is a strictly localized operation, it can be highly optimized by executing across concurrent CPU cores. The entire domain is partitioned into discrete windows of physical length $\delta$, resulting in a total of $N_{\text{win}}$ subdomains. If the number of streamwise grid points within a partitioned LRVF window is denoted as $N_{x,\text{LR}}$, implementing a grid refinement factor of $K_{r}$ yields an increased number of high-resolution grid points, expressed as $N_{x,\text{HR}} = K_{r} \times N_{x,\text{LR}}$. Accounting for the forward and inverse Fourier transform operations required to attenuate the inertial-range spectral energy via the band-stop filter, the time complexity of this phase scales as $\mathcal{O}\left(\frac{N_{\text{win}}}{P} \cdot N_z \cdot N_{x,\text{HR}} \right)$, where $P$ represents the number of parallel processors. Correspondingly, the auxiliary space complexity scales as $\mathcal{O}(N_{\text{win}} \cdot N_{z} \cdot N_{x,\text{HR}})$.

    \item \textbf{Swirling Strength $\lambda_{ci}$ and the Distribution of $\omega_{y}$:} Since locating the shear layers within the HRVF for the vortex superimposition and finding the distribution of the vorticity of the velocity field for the percentile-matching algorithm are local operations (details in Ehsani {\it et al.} \cite{ehsani2026} and the Appendix \ref{app:equation}), these phases are executed via parallel processing as well. Computing the $\lambda_{ci}$ to isolate the shear layers is vectorized, wherein the imaginary components of the complex eigenvalues are evaluated analytically utilizing the velocity gradient tensor's trace and determinant invariants. The distribution of the spanwise vorticity ($\omega_{y}$) cannot be executed within the same parallel loop as the $\lambda_{ci}$ computation; the algorithm requires the ($\lambda_{ci,\text{rms}}$) field beforehand to establish the filtering thresholds for weak shear layers. The space complexity for both operations are scales as $\mathcal{O}(N_{\text{win}} \cdot N_{z} \cdot N_{x,\text{HR}})$, whereas the time complexity are $\mathcal{O}\left(\frac{N_{\text{win}}}{P} \cdot N_z \cdot N_{x,\text{HR}} \right)$, and $\mathcal{O} \left(\frac{N_{\text{win}}}{P} \cdot \langle N_{\text{steps}} \rangle \cdot N_z \cdot N_{x,\text{HR}}\right)$ for $\lambda_{ci}$ and $\omega_{y}$, respectively.

    \item \textbf{Vortex Generation and Superimposition (SGVF-HR+VorX):} To evaluate the vorticity distributions required by the percentile-matching algorithm, the discrete vortex populations are generated prior to their structural injection into the HRVF. Because this generation step is fully vectorized and relies on inverse transform sampling, its computational overhead is negligible. The subsequent seeding phase relies on global index tracking, which introduces serial dependencies that prevent trivial parallelization across the spatial domain. Bounded conservatively by a worst-case scenario wherein a dense, highly saturated vortex population is mapped over large fractions of the grid, the execution time complexity of this seeding phase scales as $\mathcal{O}\left(N_{\text{win}} N_z N_{x,\text{HR}}^2\right)$. Crucially, because the vortex velocity fields are superimposed in-place directly onto the pre-allocated HRVF array structures, this phase requires only minimal transient memory for localized sub-grids. Consequently, the auxiliary space complexity is bounded at $\mathcal{O}(1)$, ensuring the total memory footprint of the model does not scale beyond its baseline storage requirements.

    \item \textbf{Convolution of 2-D Gaussian filter:} To simulate the viscous dissipation of the energy, a small-scale 2-D Gaussian filter is convolved into the SGVF-HR+VorX. The theoretical time complexity of this smoothing stage is bounded by $\mathcal{O}\left(N_{\text{win}} \cdot N_z \cdot N_{x,\text{HR}} \cdot K_z \cdot K_x\right)$ for a direct spatial implementation. However, because the localized Gaussian kernel size ($K_z \times K_x$) remains small and invariant relative to the global grid, and because MATLAB leverages highly optimized multi-threaded algorithms (including internal Fast Fourier Transforms where applicable), the real-world execution time is negligible.
\end{itemize}

\subsection{Limitations}
\noindent Specific structural limitations of this framework are inherent to the UMZ-based generation and concatenation process. For completeness, we summarize here the limitations of the model since its development \citep{ehsani2024stochasticprofile, ehsani2024stochasticfield, ehsani2026}.

\noindent 1) The required modeling parameters $\lambda_{T}$, $z_{0}$, $u_{\tau}$, $\delta$ are not known a priori. Scaling and empirical relationships exist to place reasonable guesses, e.g., elaborating on $\lambda_{T}=(15\nu\overline{u^{\prime 2}}/\epsilon)^{1/2}$, but rigorous estimates based on given surface geometry and undisturbed flow velocity are not strictly possible.\\
\noindent 2) The model is still two-dimensional, statistically homogeneous in the streamwise direction, and not evolving in time.\\
\noindent 3) The streamwise concatenation procedure does not privilege any preferential directionality, implying that the ramp-like structures are statistically symmetric and the structure inclination angle is flat as opposed to the $9^{\circ}-15^{\circ}$ degree angle range emerging from the literature on smooth and rough-wall \citep{adrian2000vortex, mathis2009comparison, guala_adrian10, chung2010large}.\\
\noindent 4) The remaining underestimation of the streamwise velocity energy spectrum $E_{11}$ in the dissipation subrange results from the lack of energetic sites to seed the vortices. In this current framework, they are limited to the shear layers and near-wall region. Introducing a supplementary, random background population methodology within the logarithmic regions represents a viable pathway to systematically augment high-wavenumber energy content.\\
\noindent 5) The spatial concatenation algorithm organizes the 2-D domain using a streamwise fluctuating velocity-driven ($u^{\prime}$-based) conditioning scheme, combined with the vertical spatial averaging of the wall-normal velocity within individual UMZ steps during statistics extraction \citep{ehsani2024stochasticprofile}, inducing an energy deficit in the vertical velocity component. This affects the cross-stream variance $\overline{w^{\prime 2}}/u_{\tau}^2$ and the large-scale vertical velocity spectrum $E_{22}$, a deficiency that subsequently propagates into the lower frequencies of the shear-stress cospectrum $-E_{12}$ and the Reynolds shear stress $-\overline{u^{\prime}w^{\prime}}/u_{\tau}^2$.\\
\noindent 6) A robust, generalized formulation for the model storage capacity $M$, which fundamentally governs the corresponding streamwise spacing between the concatenated 1-D velocity profiles, has not yet been mathematically closed across arbitrary Reynolds numbers and wall-normal degrees of freedom. Resolving these two limitations requires a systematic refinement of the profile concatenation methodology. 

\section{Conclusion}\label{sec:conclude}

\noindent This manuscript integrates three hierarchies of stochastic-empirical models leveraging the ubiquity of Uniform Momentum Zones and vortex cores in wall turbulence.
The generation of instantaneous velocity profiles, their subsequent spatial concatenation, and the seeding of vortices in the available shear layers have been formulated using key physical scaling variables \citep{ehsani2024stochasticprofile, ehsani2024stochasticfield, ehsani2026} and extensively validated against benchmark experimental datasets over fully rough conditions \citep{heisel2020mixing, iungo2024grand}. 
The main outcome of the current work is that these processes are reasonably described and scalable, well enough that a synthetic field can be reconstructed for an untested flow condition without requiring a supporting UMZ or vortex calibration dataset - hence no need for any PIV field. This operation was performed to successfully reproduce, comparing across various scale-dependent, second-order statistics, one of the highest Reynolds number turbulent boundary layers ever measured under laboratory conditions \citep{saddoughi1994local}. 

\noindent As discussed in the limitations section, the required parameters $z_{0}$, $u_{\tau}$, $\delta$, in addition to the estimate of $\lambda_{T}$ at the top of the roughness sublayer, have to be well guessed to correctly generate a new flow condition. This is certainly a key issue, but it also provides a deep lateral insight into the minimal parameterization of rough-wall turbulent flows in the logarithmic region that can be afforded to reproduce the near-surface atmospheric layer.

\noindent Future extensions of this framework will focus on expanding the stochastically generated velocity field from a two-dimensional to a fully three-dimensional divergence-free domain by characterizing and integrating the spanwise statistics of the uniform momentum zones. Ultimately, this scalable, structurally informed architecture is envisioned to serve as an advanced wall-modeling framework for initializing large eddy simulations, and potentially for an improved low-overhead estimate of instantaneous wall-shear stress.

\section*{Acknowledgements}
\noindent M. Guala and R. Ehsani gratefully acknowledge the support of the NSF fluid dynamics program (Award 2412025)

\section*{Code Availability}

\noindent The underlying source code for the stochastic generation of the wall turbulent boundary layer flow and statistical extraction algorithms is openly available on GitHub at \url{https://github.com/Roozbeh96/SG-of-Velocity-field-Full-Implementation}. The repository includes scripts for generating step-like UMZ profiles, smoothing shear layers, reorganization to construct a correlated 2-D velocity field, resolution enhancement, vortex generation and insertion in the shear layers and near-wall region, and spectral and spatial analysis of the generated velocity field.

\clearpage
\appendix
\section{UMZ and Vortex Instance Generation}
\label{app:equation}

Table \ref{tab:equation} outlines the analytical inverse cumulative distribution functions (c.d.f.) utilized to generate the structural parameters of the UMZs and vortex cores, while Table~\ref{tab:coeff} compiles the corresponding formulations governing the mean $(\mu)$ and the standard deviation $(\sigma)$ values of these statistical distributions.\\

\begin{table}[ht!]
    \renewcommand{\arraystretch}{2.5}
    \centering
    \caption{\label{tab:equation} Inverse c.d.f. for the generation of a UMZ's characteristics and vortex attributes (thickness $h_{m_{\mathnormal{i}}}$, modal $u_{m_{\mathnormal{i}}}$ and vertical velocity $w_{m_{\mathnormal{i}}}$, radius $r_{\omega_{i}}$, maximum azimuthal velocity $u_{\omega_{i}}$, and Pearson correlation coefficient $\rho_{\omega_{j}}^{u,w}$ used to transform streamwise and vertical velocity component of a generated vortex). The red line separates the governing equations used for UMZ generation from those implemented for Vortex generation.}
    \begin{tabularx}{\textwidth}{c|Y}
        \textbf{Variables} & \textbf{Inverse c.d.f.} \\ \hline
        
        $h_{m_{\mathnormal{i}}}$ & \text{Log-normal}:\;$z_{\mathnormal{i}}\exp(\hat{\mu}_{_{H_{m}}}^{z_i}+\hat{\sigma}_{_{H_{m}}}^{z_i}\sqrt{2}\;\text{erfinv}(2\;\text{rand}_{h_{m_{\mathnormal{i}}}}-1)))$ \\ \hline
        
        $u_{m_{\mathnormal{i}}}$ & \text{Gaussian}:\;$u_\tau(\mu_{_{U_{m}}}^++\sigma_{_{U_{m}}}^+\sqrt{2}\;\text{erfinv}(2\;\text{rand}_{u_{m_{\mathnormal{i}}}}-1))$ \\ \hline
        
        $w_{m_{\mathnormal{i}}}$ & \text{Gaussian}:\;$u_\tau(\mu_{_{W_{m}}}^++\sigma_{_{W_{m}}}^+\sqrt{2}\;\text{erfinv}(2\;\text{rand}_{w_{m_{\mathnormal{i}}}}-1))$ \\ \arrayrulecolor{red}\hline \arrayrulecolor{black}
        $r_{\omega_{j}}$ & 
        $ \begin{array}{l l}
            \text{Log-normal}:\lambda_{T}\exp(\mu_{r_{\omega}}+\sigma_{r_{\omega}}\text{norminv}(\text{rand}_{r_{\omega_{j}}})) & \text{rand}_{r_{\omega_{j}}}<\text{rand}_{x_{t}} \\
            \text{Pareto}:\lambda_{T}x_{t}(\frac{1-\text{rand}_{r_{\omega_{j}}}}{1- \text{rand}_{x_{t}}})^{-1/\alpha} & \text{rand}_{r_{\omega_{j}}}\geq \text{rand}_{x_{t}} 
        \end{array} $ \\ \hline
        
        $u_{\omega_{j}}$ &
        {\text{Exponential}:\;$ -u_{\tau}m\ln(-\text{rand}_{u_{\omega_{j}}}(\exp(-a/m)-\exp(-b/m))+\exp(-a/m)) $} \\ \hline
        
        $\rho_{\omega_{j}}^{u,w}$ & 
        {\text{Skewed Gaussian}:\;$\text{inv}(\int_{-1}^{\rho_{\omega}^{u,w}} C_{1} \left(1 - \left(\rho_{\omega}^{u,w}\right)^2 \right) \frac{2}{\omega} \phi \left( \frac{\rho_{\omega}^{u,w} - \xi}{\omega} \right) \mathrm{\Phi} \left( \beta \left( \frac{\rho_{\omega}^{u,w} - \xi}{\omega} \right) \right) d\rho_{\omega}^{u,w})$
        
        $-1 \le \rho_{\omega}^{u,w} \le 1, \quad \phi: \text{normpdf}, \quad \mathrm{\Phi}: \text{normcdf}$} \\
    \end{tabularx}
\end{table}

\begin{table}[ht!]
    \renewcommand{\arraystretch}{2.0}
    \centering
    \caption{\label{tab:coeff} Statistical parameters governing the mean $(\mu)$ and the standard deviation $(\sigma)$ of the inverse c.d.f. equations utilized for UMZ characteristics and vortex features generation. The red line separates the parameters used for UMZ generation from those used for Vortex generation. For vortex generation, depending on whether the vortex center location is $z_{j,c} < 2k_{s}$ or $z_{j,c} \ge 2k_{s}$, different parameters will be used.}
    \resizebox{\textwidth}{!}{ 
    \begin{tabular}{c|c}
        \textbf{Variables} & \textbf{Parameters of the Inverse c.d.f. Equations} \\
        \hline
        $h_{m_{\mathnormal{i}}}$ & $\hat{\mu}_{_{H_{m}}}^{z_i}\simeq-3.59(\frac{z_{i}}{\delta})^{0.91}, \hat{\sigma}_{_{H_{m}}}^{z_i} \simeq 1$\\
        \hline
        $u_{m_{\mathnormal{i}}}$ & $ 
        \mu_{_{U_{m}}}^{+} = \frac{1}{\kappa} \ln{(\frac{z_{i}+0.5h_{m_{i}}}{z_0})}, \sigma_{_{U_{m}}}^{+} \simeq \sqrt{-A\ln{(\frac{z_{i}+0.5h_{m_{i}}}{\delta})}+B}\quad A = 1.26, B = 1$\\
        \hline
        $w_{m_{\mathnormal{i}}}$ & $\mu_{_{W_{m}}}^{+}=0, \sigma_{_{W_{m}}}^{+} \simeq 0.85$\\
        \arrayrulecolor{red}\hline \arrayrulecolor{black}
        $r_{\omega_{j}}$ for $z_{j,c}/k_{s} < 2$ & $\mu_{r_{\omega}} = -1.55, \sigma_{r_{\omega}} = 0.36, x_{t} = 0.36, \text{rand}_{x_{t}} = 0.93, \alpha = 5.0$ \\
        
        $r_{\omega_{j}}$ for $z_{j,c}/k_{s} \geq 2$ & $\mu_{r_{\omega}} = -1.94, \sigma_{r_{\omega}} = 0.36, x_{t} = 0.25, \text{rand}_{x_{t}} = 0.93, \alpha = 4.5$ \\
        \hline
        $u_{\omega_{j}}$ for $z_{j,c}/k_{s} < 2$ & $m = 0.55, a = 0.65, b = 4.5$ \\
        $u_{\omega_{j}}$ for $z_{j,c}/k_{s} \geq 2$ & $m = 0.44, a = 0.4, b = 4.5$ \\
        \hline
        $\rho_{\omega_{j}}^{u,w}$ for $z_{j,c}/k_{s} < 2$ & $C_{1} = 1.23, \omega = 0.58, \xi = -0.61, \beta = 2.37$ \\
        $\rho_{\omega_{j}}^{u,w}$ for $z_{j,c}/k_{s} \geq 2$ & $C_{1} = 1.28, \omega = 0.66, \xi = -0.54, \beta = 1.80$ \\
    \end{tabular}
    }
\end{table}

\noindent \textit{UMZ Generation}: The algorithmic synthesis of an individual instantaneous velocity profile initiates at a baseline wall-normal location of $z_{i} = z_{\text{start}} = 1.5k_{s}$. Using a random number sampled from a uniform distribution, the $i$-th UMZ's characteristics are generated. For the thickness, we use the statistics of $z_{i}$; however, for modal and vertical velocity, we use statistics of $z_{i,c} = z_{i}+h_{m_{i}}/2$, where $z_{i,c}$ is the midpoint of the generated UMZ (shown in Table \ref{tab:coeff}, rows 2 and 3). The modeling terms $\hat{\mu}_{H_{m}}^{z_i}$ and $\hat{\sigma}_{H_{m}}^{z_i}$ represent the mean and standard deviation of the logarithm of UMZs' thicknesses normalized by $z_{i}$, both of which are modeled (i.e., $\hat{\mu}_{H_{m}}^{z_i} = \mu(\log{(h_{m_{i}}}/z_{i}))$, $\hat{\sigma}_{H_{m}}^{z_i} = \sigma(\log{(h_{m_{i}}}/z_{i}))$ ). Upon generating $h_{m_{i}}$, the interface bound is advanced to the next upward wall-normal position, $z_{i+1} = z_{i}+h_{m_{i}}$, and the sampling loop is repeated iteratively to generate the UMZs' characteristics and stack them on top of each other. The process of generating one profile ends when the $z_{i+n} > z_{\text{end}}$ where $z_{\text{end}} = 0.25\delta$. Then all the realizations beyond $z_{\text{end}}$ will be truncated, resulting in a profile that starts from $z_{\text{start}}$ and extends to $z_{\text{end}}$. Log-normal and Gaussian distributions are used to generate thickness, modal, and vertical velocities of an UMZ instance, respectively.\\

\noindent \textit{Vortex Generation}: The Oseen vortex model \cite{oseen1912uber} is utilized to formulate the velocity field of the $j$-th injected vortex instance, which is subsequently superimposed onto the HRVF. 
To account for the sharp structural transition from the highly anisotropic near-wall state to the more isotropic state characteristic of outer-layer vortices, distinct parameterizations are prescribed based on the wall-normal core coordinate $z_{j,c}$. This structural shift directly modulates the mean of the cross-component Pearson correlation coefficient distribution ($\rho_{\omega_{j}}^{u,w}$), distinguishing cores situated within the roughness sublayer ($z_{j,c} < 2k_{s}$) from those residing in the outer log-layer domain ($z_{j,c} \ge 2k_{s}$).
Furthermore, the physical domain bounds of the vortex velocity fields are constrained: the truncation radius for shear-layer structures extends to $r_{\text{max}} = 2 r_{\omega_{j}}$, whereas near-wall forcing vortices are truncated at $r_{\text{max}} = r_{\omega_{j}}$. 
To ensure physical compatibility between the velocity field and the generated vortex, a percentile-matching algorithm adapts the vorticity of each injected core to match the localized vorticity of the background shear layers at the centroid.
The fundamental attributes, core radius $r_{\omega_{j}}$, maximum azimuthal velocity $u_{\omega_{j}}$, and the structural correlation coefficient $\rho_{\omega_{j}}^{u,w}$, are modeled using piece-wise lognormal–Pareto, exponential, and skewed Gaussian distributions, respectively. The core radius is generated using a conditional probability threshold: a log-normal formulation is evaluated if the random number satisfies $\text{rand}_{r_{\omega_{j}}} < 0.93$, whereas a heavy-tailed Pareto distribution is implemented for the upper tail regime where $\text{rand}_{r_{\omega_{j}}} \ge 0.93$. 
 Because the skewed Gaussian distribution governing $\rho_{\omega_{j}}^{u,w}$ lacks a closed-form analytical inverse, its cumulative probability density function is integrated numerically and subsequently mapped via a discrete inversion scheme. 
 A detailed explanation about the generation of primary, secondary, and near-wall vortices, and the percentile-matching algorithm is provided in Ehsani {\it et al.}\cite{ehsani2026}.

\bibliographystyle{unsrtnat}
\bibliography{Ref}

\end{document}